\pdfoutput=1
\documentclass[ALICE,manyauthors]{cernphprep}

\usepackage[comma,square,numbers,sort&compress]{natbib}
\usepackage{hyperref}
\usepackage{lineno}

\usepackage[utf8]{inputenc}
\usepackage{amsmath}
\usepackage{amsfonts}
\usepackage{amssymb}
\usepackage{amsthm}
\usepackage{bm}
\usepackage[english]{babel}
\usepackage{fontenc}
\usepackage{graphicx}
\usepackage{enumerate}
\usepackage{textcomp}
\usepackage[toc,page]{appendix}
\usepackage{slashed}
\usepackage{lineno}
\usepackage{color}

\usepackage{setspace}

\graphicspath{
 {IMAGES/}
}

\newcommand{ \be }{\begin{linenomath*}\begin{eqnarray}}
\newcommand{ \ee }{\end{eqnarray}\end{linenomath*}}
\newcommand{ \la }{\langle}

\newcommand{ \ra }{\rangle}

\newcommand{ \pt }{p_{\rm T}}

\newcommand{\PbPb}{Pb--Pb }
\newcommand{\etas}{\eta/s}

\newcommand{\GeVc}         {Ge\kern-.1emV/$c$\xspace}
\newcommand{\MeVc}         {Me\kern-.1emV/$c$\xspace}
\newcommand{\TeVe}         {Te\kern-.1emV\xspace}
\newcommand{\GeVe}         {Ge\kern-.1emV\xspace}
\newcommand{\MeVe}         {Me\kern-.1emV\xspace}
\newcommand{\GeVmass}      {Ge\kern-.1emV/$c^2$\xspace}
\newcommand{\MeVmass}      {Me\kern-.1emV/$c^2$\xspace}

\definecolor{dgreen}{cmyk}{1.,0.,1.,0.2}        
\definecolor{orange}{cmyk}{0.,0.353,1.,0.}    

\begin{document}%

\begin{titlepage}
\PHyear{2019}
\PHnumber{252}      
\PHdate{28 October}  
%

\title{Longitudinal and azimuthal evolution of\\two-particle transverse momentum correlations\\in Pb--Pb collisions at $\mathbf{\sqrt{\text{\textit{s}}_{\text{NN}}} = 2.76\;\text{\TeVe}}$}
\ShortTitle{$G_{2}$ transverse momentum correlation}   

\Collaboration{ALICE Collaboration\thanks{See Appendix~\ref{app:collab} for the list of collaboration members}}
\ShortAuthor{ALICE Collaboration} 

\begin{abstract} 
This paper presents the first measurements of the charge independent (CI) and charge dependent (CD)  two-particle transverse momentum correlators $G_{2}^{\rm CI}$ and $G_{2}^{\rm CD}$  in Pb--Pb collisions at $\sqrt{s_{\text{NN}}} = 2.76\;\text{\TeVe}$ by the ALICE collaboration. The two-particle transverse momentum correlator $G_{2}$ was introduced as a measure of the momentum current transfer between  neighbouring system cells.
The correlators are measured as a function of pair separation in pseudorapidity ($\Delta \eta$) and azimuth ($\Delta \varphi$) and as a function of collision centrality.
From peripheral to central collisions, the correlator $G_{2}^{\rm CI}$ exhibits a longitudinal broadening while undergoing a monotonic azimuthal narrowing. By contrast, $G_{2}^{\rm CD}$ exhibits a narrowing along both dimensions. 
These features are not reproduced by models such as HIJING and AMPT. However, the observed narrowing of the correlators from peripheral to central collisions is expected to result from the stronger transverse flow profiles produced in more central collisions and  the longitudinal broadening is predicted to be sensitive  to momentum currents and the shear viscosity per unit of entropy density $\eta/s$ of the matter produced in the collisions. The observed broadening is found to be consistent with the hypothesized lower bound of $\eta/s$ and is in qualitative agreement with values obtained from anisotropic flow measurements. 

\end{abstract}
\end{titlepage}
\setcounter{page}{2}

\section{Introduction}
Measurements of particle production and their correlations performed at the Relativistic Heavy Ion Collider (RHIC) and the Large Hadron Collider (LHC) provide compelling evidence that the matter produced in heavy-ion collisions is characterized by extremely high temperatures and energy densities consistent with a deconfined, but strongly interacting Quark--Gluon Plasma (QGP)~\cite{Adams:2005dq,Adcox:2004mh,Arsene20051,Back:2004je}. 
Collective flow, which manifests itself by the anisotropy of particle production in the plane transverse to the beam direction, is  characterized by the harmonic coefficients of a Fourier expansion of the azimuthal distribution of particles relative to the reaction plane. Comparisons of these harmonic coefficients with hydrodynamical model predictions indicate that the matter produced in those collisions has a shear viscosity per unit of entropy density, $\etas$, that nearly vanishes~\cite{Adcox:2004mh,Heinz:2011kt}. 
The shear viscosity quantifies the resistance that any medium presents to its anisotropic deformation. It contributes to the transfer of momentum from one fluid cell to its neighbors as well as  the damping of momentum fluctuations. The reach of $\eta/s$ effects is expected to grow with the lifetime of the system.
Recent measurements of flow coefficients  and  hydrodynamical predictions largely focus on the precise determination of  $\etas$~\cite{Adams:2004bi,Aamodt:2011by,Heinz:2013th,PhysRevLett.105.252302}. However, quantitative descriptions of heavy-ion collisions with hydrodynamical models generally rely on specific parametrizations of  the initial conditions of colliding systems, i.e., their initial energy and entropy density distribution in the transverse plane, the magnitude of initial fluctuations, the thermalization time, and several model parameters.  
It is found that the precision of model predictions is  hindered, in particular, by uncertainties in the initial state conditions. Indeed, values of shear viscosity that best match the observed flow coefficients are dependent on the initial conditions, and unless the magnitude of the initial state fluctuations can be precisely assessed, the achievable precision on $\etas$ might remain limited~\cite{Song:2010mg,Shen:2012vn}. 
Systematic studies of correlations between different order harmonic coefficients~\cite{Acharya:2017gsw}, shown to be sensitive to the initial conditions and the temperature dependence of $\etas$, can help to provide further constraints to those conditions and to the transport properties of the system.
Novel approaches based on Bayesian parameter estimation~\cite{PhysRevC.97.044905,Bernhard:2016tnd} bring progress on a simultaneous characterization of the initial conditions and the QGP. Furthermore, it was pointed out~\cite{Gavin:2006xd} that the strength of momentum current correlations may be sensitive to $\etas$. It was shown, in particular, that the longitudinal broadening of a transverse momentum ($\pt$) correlator, formally defined below and hereafter named $G_2$,  with increasing system lifetime is directly sensitive to $\etas$ while it does not have any explicit dependence on the initial state fluctuations in the transverse plane of the system. 

A first measurement of the broadening of the two-particle transverse momentum correlator $G_2$ was reported by the STAR collaboration~\cite{Agakishiev:2011fs}. 
Improved techniques to correct  for instrumental effects have since then been reported~\cite{Ravan:2013lwa,Acharya:2018ddg,Gonzalez:2018cty}. 
In this letter, these techniques are used to measure differential charge independent (CI) and charge dependent (CD) two-particle transverse momentum correlators, $G_{2}^{\rm CI}$ and $G_{2}^{\rm CD}$, respectively, as a function of pair rapidity difference, $\Delta\eta$,  and azimuthal angle difference, $\Delta\varphi$,  for selected ranges of  \PbPb collision centrality.
The shapes of these correlators are studied with a two-component model and the longitudinal and azimuthal widths of their near-side peaks are studied as a function of the \PbPb collision centrality. The longitudinal broadening  of  $G_{2}^{\rm CI}$ from peripheral to central collisions is used to assess the magnitude of $\eta/s$ of the matter produced in \PbPb collisions while the longitudinal and azimuthal widths of $G_{2}^{\rm CD}$ are used to assess the role of competing effects, including  radial flow, diffusion, and the broadening of jets  by interactions with the medium. 
In that context, measurements of $G_{2}$ are also compared with previously reported measurements of the two-particle number correlator $R_2$ and two-particle transverse momentum correlator $P_2$~\cite{Acharya:2018ddg}.

\section{\texorpdfstring{The $G_{2}$ correlator}{}}
The dimensionless variant of the $G_2$ correlator~\cite{Gavin:2006xd,Sharma:2008qr} reported in this letter is  defined according to 
\begin{equation}
G_2\left( \eta_1,\varphi_1,\eta_2,\varphi_2 \right) =  
  \frac{1}{\langle p_{\rm T,1} \rangle \langle p_{\rm T,2} \rangle} \left[ 
  \frac{\int_{\Omega} p_{\rm T, 1}p_{\rm T, 2}\, \rho_2(\vec p_1,\vec p_2) 
      \, {\rm d}\,p_{\rm T, 1}{\rm d}\,p_{\rm T, 2}}
    {\int_{\Omega} \rho_1(\vec p_1)\, {\rm d}\,p_{\rm T, 1}
      \; \int_{\Omega} \rho_1(\vec p_2)\, {\rm d}\,p_{\rm T, 2}} 
    - \la p_{\rm T,1}\ra(\eta_1,\varphi_1) \la p_{\rm T,2}\ra(\eta_2,\varphi_2)
    \right]
  \label{eq:normG2}
\end{equation}
 where $\Omega$ is the phase space region in which the measurement is performed; $\vec{p}_{1}$ and $\vec{p}_{2}$ are the three-momentum vectors of particles of a given pair;  $p_{{\rm T},1}$ and $p_{{\rm T},2}$ their  transverse momentum components, respectively; $\rho_{1}(\vec{p}_{i}) = {\rm d}^{3}N/{\rm d}p_{{\rm T},i}\,{\rm d}\eta_{i}\,{\rm d}\varphi_{i}$ and $\rho_{2}(\vec{p}_{1},\vec{p}_{2}) = {\rm d}^{6}N/{\rm d}p_{{\rm T},1}\,{\rm d}\eta_{1}\,{\rm d}\varphi_{1}\,{\rm d}p_{{\rm T},2}\,{\rm d}\eta_{2}\,{\rm d}\varphi_{2}$ represent single and pair particle densities, expressed as functions of $\vec{p}_{i}$, $i=1,2$, and $(\vec{p}_{1},\vec{p}_{2})$, respectively; $\la p_{\rm T}\ra(\eta_i,\varphi_i)$ is the average  transverse momentum of particles observed at $(\eta_i,\varphi_i)$, with $\eta_i,\varphi_i$, $i=1,2$, referring to single-track pseudorapidity and azimuthal angle, respectively; and $\langle p_{{\rm T},i} \rangle = \int \rho_{1}(\vec{p}_{i})\, p_{{\rm T},i} \, {\rm d}\vec{p}_{i}$ is the inclusive average transverse momentum of produced particles, $i = 1,2$, in the considered event ensemble. Experimentally, $G_2$ is 
calculated as 
\begin{equation}
  {G}_2 \left(\eta_1,\varphi_1,\eta_2,\varphi_2 \right)  
    = \frac{1}{\langle p_{\rm T,1} \rangle \langle p_{\rm T,2} \rangle} \left[
      \frac{S_{p_{\text{T}}}(\eta_1,\varphi_1,\eta_2,\varphi_2) }{\langle n_{1,1}(\eta_1,\varphi_1) \rangle \langle n_{1,2}(\eta_2,\varphi_2) \rangle} 
        - \la p_{\rm T,1}\ra(\eta_1,\varphi_1) \la p_{\rm T,2}\ra(\eta_2,\varphi_2) 
    \right]
  \label{eq:gavinG}
\end{equation}
with
\begin{equation}
S_{p_{\text{T}}}(\eta_1,\varphi_1,\eta_2,\varphi_2) =  \left\langle \sum\limits_{\text{i}}^{n_{1,1}} \sum\limits_{\text{j} \neq \text{i}}^{n_{1,2}} 
        p_{\text{T},\text{i}} \; p_{\text{T},\text{j}}\right\rangle
\end{equation}
where $n_{1,1}$ and $n_{1,2}$ are the number of tracks on each event within bins centered  at $\eta_1,\varphi_1$ and $\eta_2,\varphi_2$, and with transverse momentum $p_{{\rm T},i}$, $i \in [1,n_{1,1}]$, and $p_{{\rm T},j}$, $j \neq i \in [1,n_{1,2}]$, respectively. Angle brackets, $\la \cdots \ra$, refer to event ensemble averages, $\la A \ra = \sum_{1}^{N_{\text{events}}} A / N_{\text{events}}$. 
The correlators  ${G}_2^{\rm LS}$ and ${G}_2^{\rm US}$  are first measured for like-sign (LS) and unlike-sign (US) pairs separately, and combined to obtain CI and CD correlators  according to $G_2^{\rm{CI}} = \frac{1}{2} \left(G_2^{\rm{US}}+ G_2^{\rm{LS}}\right) $ and $G_2^{\rm{CD}} = \frac{1}{2} \left(G_2^{\rm{US}} - G_2^{\rm{LS}}\right) $, respectively~\cite{Acharya:2018ddg}. 
Measurements of $G_2(\eta_1,\varphi_1,\eta_2,\varphi_2)$ are averaged across the longitudinal and azimuthal acceptances in which the measurement is performed to obtain $G_2(\Delta \eta, \Delta \varphi)$, where $\Delta \eta = \eta_1-\eta_2$ and $\Delta \varphi = \varphi_1-\varphi_2$, with a procedure similar to that used for $R_2$ and $P_2$ correlators~\cite{Acharya:2018ddg}.


\section{Measurement techniques}
The results presented in this letter are based on  $1.1\times10^{7}$ selected minimum bias (MB) \PbPb\ collisions  at $\sqrt{s_{\rm NN}}$ = 2.76 \TeVe collected during the 2010 LHC  heavy-ion run by the ALICE experiment. Detailed descriptions of the ALICE detectors and their respective performances are given in Refs.~\cite{1748-0221-3-08-S08002,Abelev:2014ffa}.
The MB trigger was configured in order to have high efficiency for hadronic events, requiring at least two out of the following three conditions: i) two hits in the second inner layer of the Inner Tracking System (ITS), ii) a signal in the V0A detector, iii) a signal in the V0C detector. The amplitudes measured in the V0 detectors are additionally used to estimate the collision centrality  reported in nine  classes corresponding to 0--5\% (most central), 5--10\%, 10--20\%, ...,  70--80\% (most peripheral)  of the total interaction cross section~\cite{Abelev:2013qoq}.  The vertex position of each collision is determined with tracks reconstructed in the ITS and the Time Projection Chamber~(TPC) and is required to be in  the range $|z_{\rm vtx}| \le 7$ cm of the nominal interaction point (IP).   Pile-up events, identified as events having multiple reconstructed vertices in the ITS, are rejected. Additionally, the extra activity observed in slow response detectors (e.g., TPC) relative to  that measured in fast detectors (e.g., V0) for out of bunch pile-up events is used to discard these events. 
The remaining event pile-up contamination is estimated to be negligible. Longitudinally, the ITS covers $|\eta|<0.9$, the TPC $|\eta|<0.9$, V0A $2.8 < \eta < 5.1$ and V0C $-3.7 <\eta < -1.7$. These four detectors feature full azimuthal coverage.

The present measurement of the $G_2$ correlators is based on charged particle tracks measured with the TPC detector in the transverse momentum  range $0.2\le \pt \le 2.0$ \GeVc and the pseudorapidity range $|\eta|<0.8$. In order to ensure good track quality and to minimize secondary track contamination, the analysis is  restricted to charged particle tracks involving a minimum of 50 reconstructed TPC space points out of a maximum of 159, and  distances of closest approach (DCA) to the reconstructed primary vertex  of less than $3.2\;\text{cm}$ and $2.4\;\text{cm}$ in the longitudinal and radial directions, respectively. 
An alternative criterion, used in the analysis of the systematic uncertainties, that relies on tracks reconstructed with the combination of the TPC and the ITS detectors, henceforth called ``global tracks'', involves a minimum of 70 reconstructed TPC space points, hits either on any of two inner layers of the ITS, or in the third inner layer of the ITS, and a tighter DCA selection criterion in both, longitudinal and radial directions, the latter one $\pt$-dependent.
Electrons (positrons), whose one of the largest sources are photon conversions into $e^+e^-$ pairs, are suppressed discarding $e^+$ and $e^-$ by removing tracks with a specific energy loss ${\rm d}E/{\rm d}x$ in the TPC closer than $3\sigma_{{\rm d}E/{\rm d}x}$ to the expected median for electrons and at least $5\sigma_{{\rm d}E/{\rm d}x}$ away from the $\pi$, $K$ and $p$ expectation values.

The single and pair efficiencies of the selected charged particles  are estimated from a Monte Carlo (MC) simulation using the HIJING event generator~\cite{Wang:1991hta} with particle transport through the detector performed with GEANT3~\cite{Brun:1082634} tuned to reproduce the detector conditions during the 2010 run.
Corrections for single track losses due to non-uniform acceptance (NUA) are carried out using a weighting technique~\cite{Ravan:2013lwa} separately for data and for reconstructed MC data. Weights are extracted separately for positive and negative tracks, for each collision centrality range,  as a function of  $\eta$, $\varphi$, $\pt$ and the longitudinal position of the primary vertex of each event, $z_{\text{vtx}}$. The $\pt$-dependent single track efficiency correction is extracted as the inverse of the ratio of the number of NUA corrected reconstructed HIJING tracks to generated tracks. Data are subsequently corrected with NUA and single track efficiency corrections.
Pair  losses due to track merging or crossing  are corrected in 
part based on the technique described in ~\cite{Acharya:2018ddg} and in part based on the ratio of the average number of reconstructed HIJING pairs relative to  the generated number of pairs.   Corrections  for  $p_{\rm T}$ dependent  pair losses  are  not included in the reported results given they have a large ($>20\%$) systematic uncertainty. Correlator values at $|\Delta \eta| < 0.05$, $|\Delta \varphi| < 0.04\;\text{rad.}$, left under-corrected by this last fact, are  not reported in this work. 
However, this does not impact the shape and width of the $G_2$ correlator, which are of interest for the determination of the viscous broadening.  No filters are used to suppress like-sign (LS) particle correlations resulting from Hanbury Brown and Twiss (HBT) effects. For pions, which dominate the particle production, HBT produces a peak centered at $\Delta \eta$, $\Delta \varphi=0 $  in $G_2^{\rm LS}$. The width of this peak decreases in inverse proportion to  the size of the collision system.  Given the number of HBT pairs is relatively small compared to the total number of pairs accounted for in $G_2^{\rm LS}$, the implied reduction of the longitudinal broadening is relatively modest and thus not considered in this analysis. 

\section{Statistical and systematic uncertainties}
Statistical uncertainties on the strength of $G_2$ are extracted using the sub-sample method with ten sub-samples. Systematic uncertainties are determined by repeating the analysis under different event and track selection conditions. 
Deviations from the nominal results are considered significant and assessed as systematic uncertainties based on a statistical test~\cite{Barlow:2002yb}. 
The impact of potential TPC effects sensitive to the magnetic field polarity is assessed by splitting the whole data sample into positive and negative magnetic field configurations, whereas uncertainties associated with the collision  centrality estimation are studied by comparing nominal results, based on the V0 detector,  with those obtained with an alternative centrality measure based on hit multiplicity on the two inner layers of the ITS. 
Effects of the kinematic acceptance in which the measurement is performed are investigated by repeating the analysis with  events in the range $|z_{\rm vtx}| < 3$ cm of the nominal IP.  The presence of  biases caused by secondary particles is checked using the ``global tracks" selection criterion. 
Biases associated  with pair losses are studied based on  pair efficiency corrections  obtained with HIJING/GEANT3 simulations.   
The largest systematic uncertainty amounts to a  global shift in $G_2(\Delta\eta,\Delta\varphi)$ correlator strength which is independent of $\Delta\eta$ and $\Delta\varphi$ and is reported as $\delta B$. This  shift affects the magnitude of the projections onto $\Delta\eta$ and $\Delta\varphi$ but not the shapes of the near-side peak, $|\Delta \varphi| < \pi/2$, of $G_2$ along these coordinates. Systematic uncertainties in the shape of the near-side peak of $G_2^{\text{CI}}$ and $G_2^{\text{CD}}$ are mainly due to the presence of secondary particles. Overall, systematic uncertainties on the shapes of the projections of $G_2^{\text{CI}}$ and $G_2^{\text{CD}}$ along the longitudinal (azimuthal) dimension amount to 4\%(5\%) and 5\%(10\%), respectively, with decreasing values towards peripheral events. 

\section{Results}
\begin{figure}[t]
  \centering
  \includegraphics[scale=0.29,keepaspectratio=true,clip=true,trim=2pt 4pt 43pt 2pt]
  {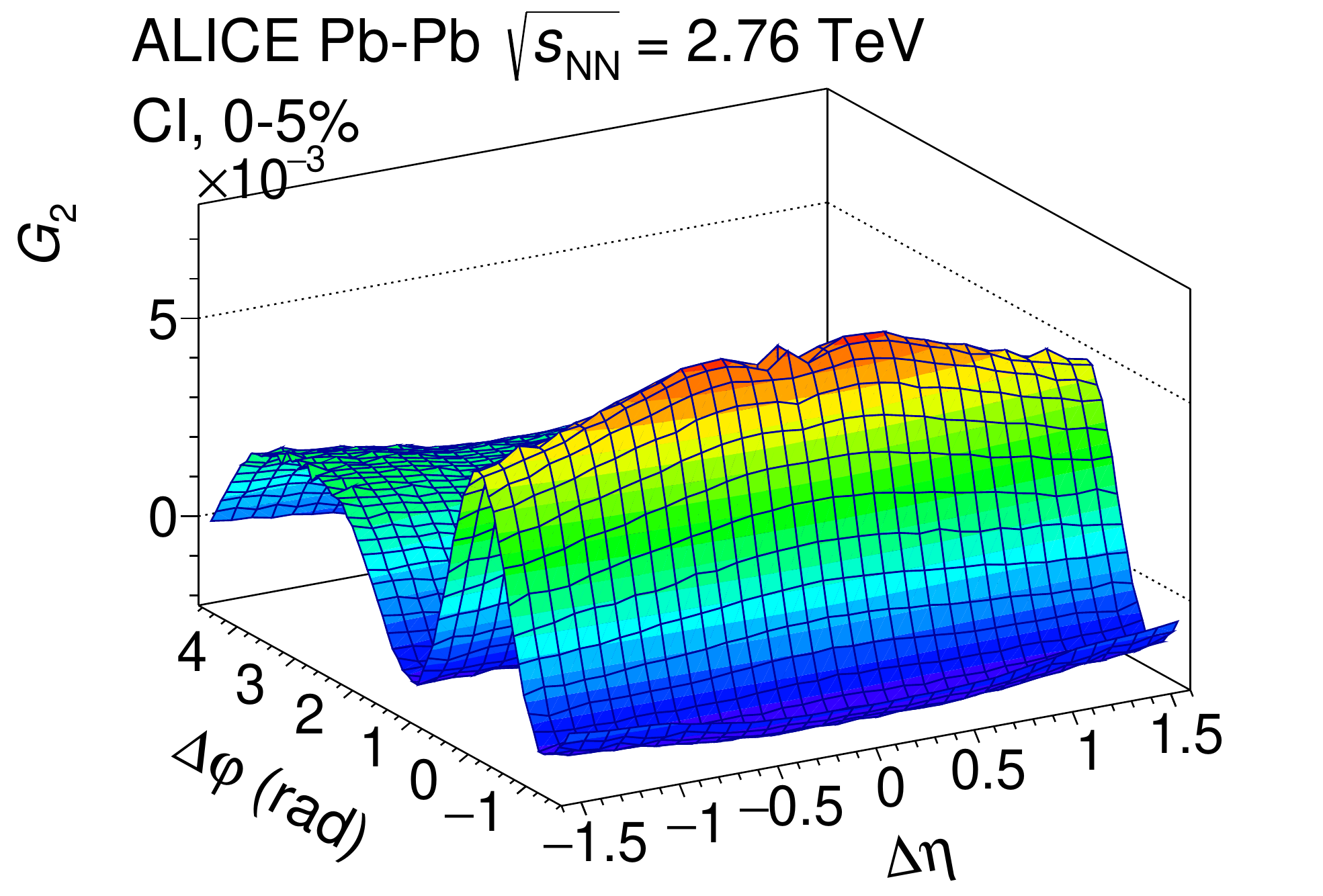}
  \includegraphics[scale=0.29,keepaspectratio=true,clip=true,trim=37pt 4pt 43pt 2pt]
  {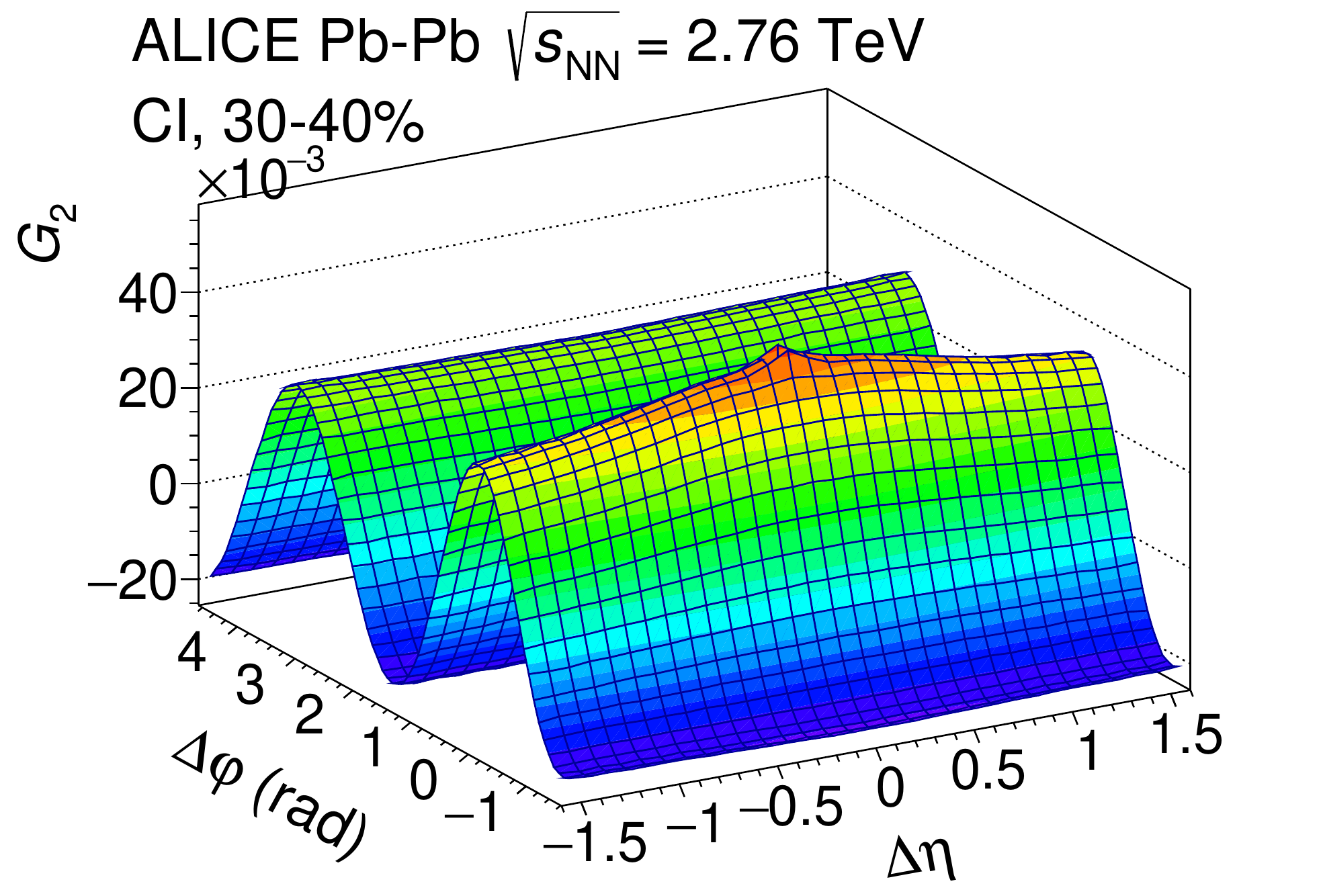}
  \includegraphics[scale=0.29,keepaspectratio=true,clip=true,trim=37pt 4pt 43pt 2pt]
  {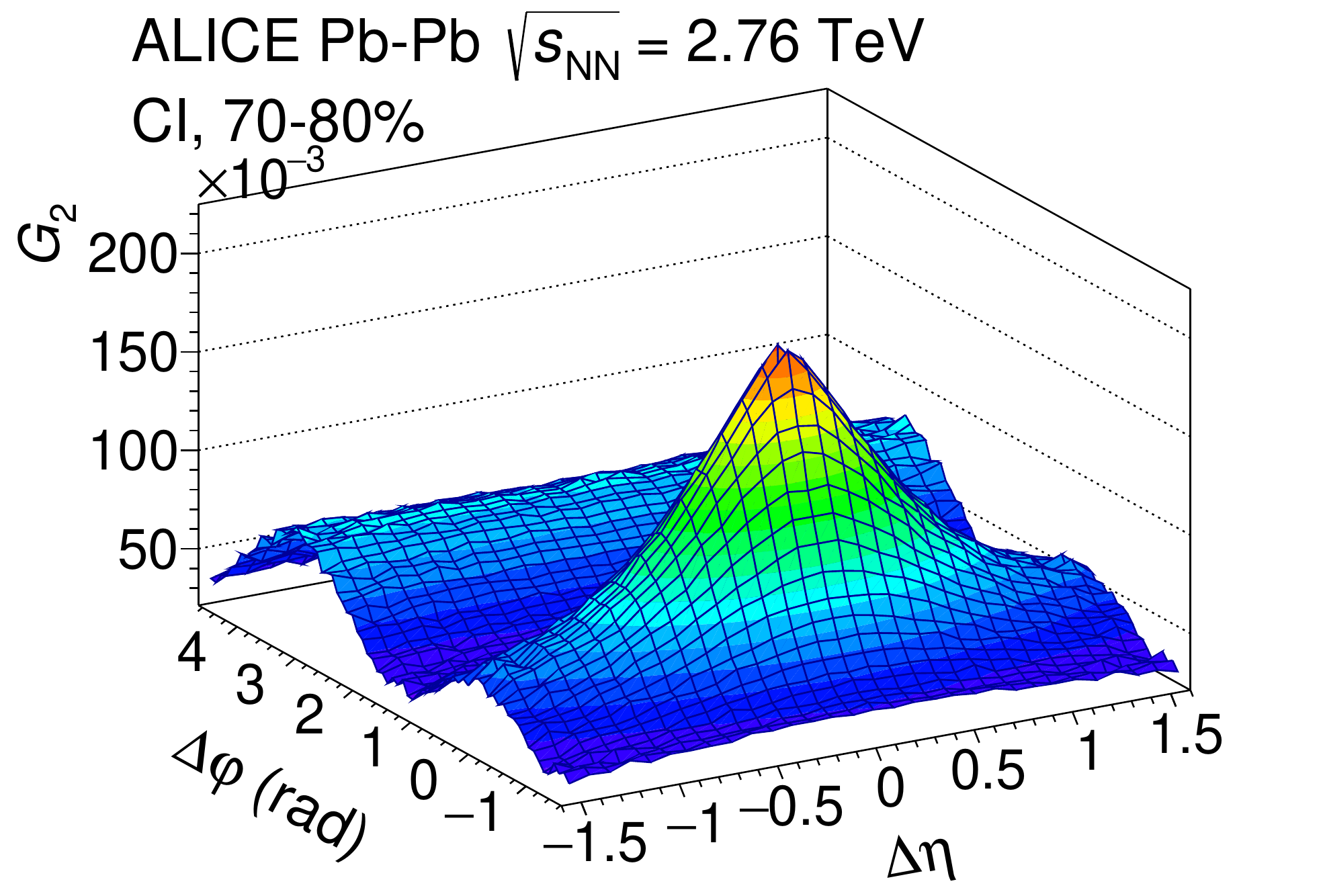} \par
  \includegraphics[scale=0.29,keepaspectratio=true,clip=true,trim=2pt 4pt 43pt 2pt]
  {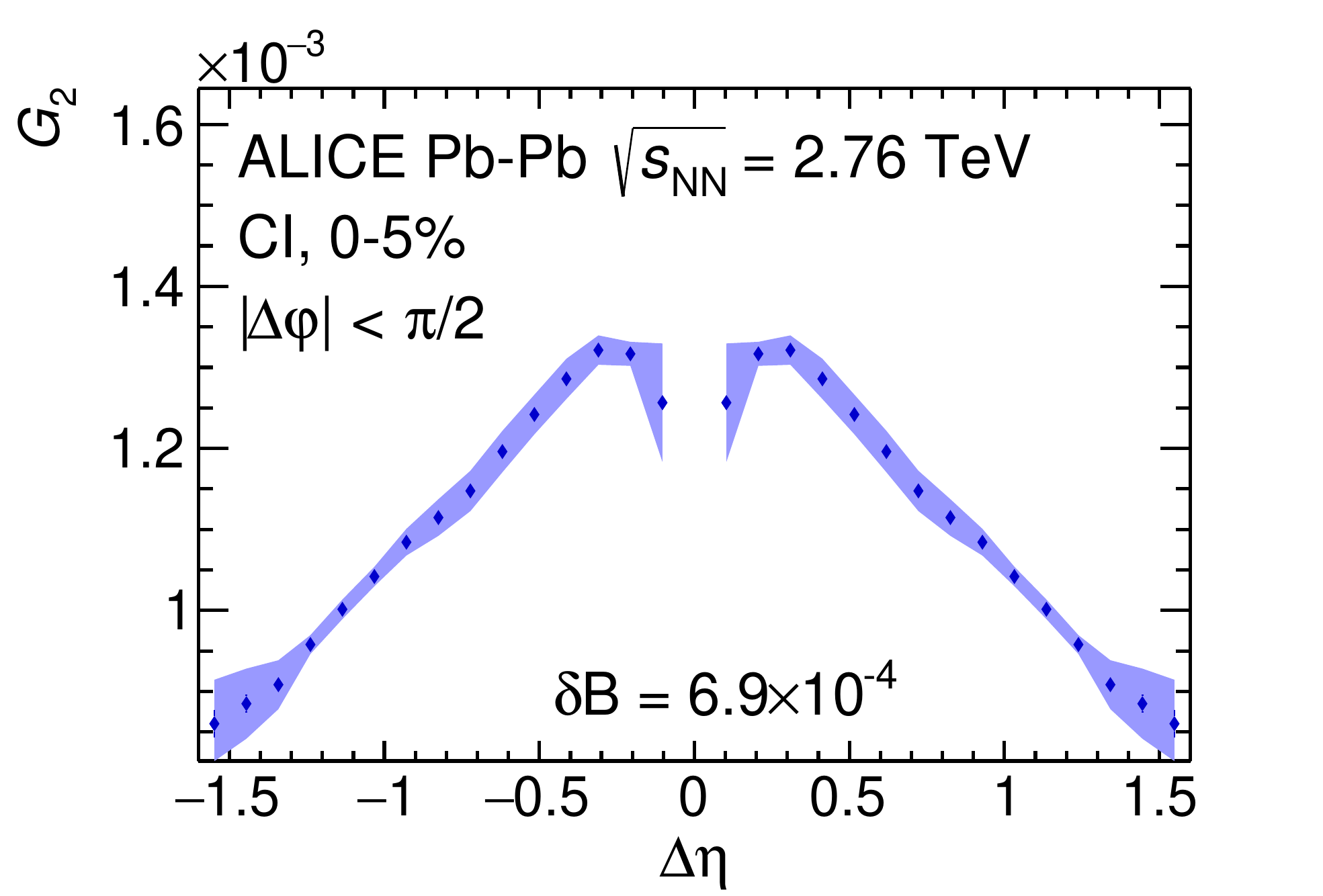}
  \includegraphics[scale=0.29,keepaspectratio=true,clip=true,trim=37pt 4pt 43pt 2pt]
  {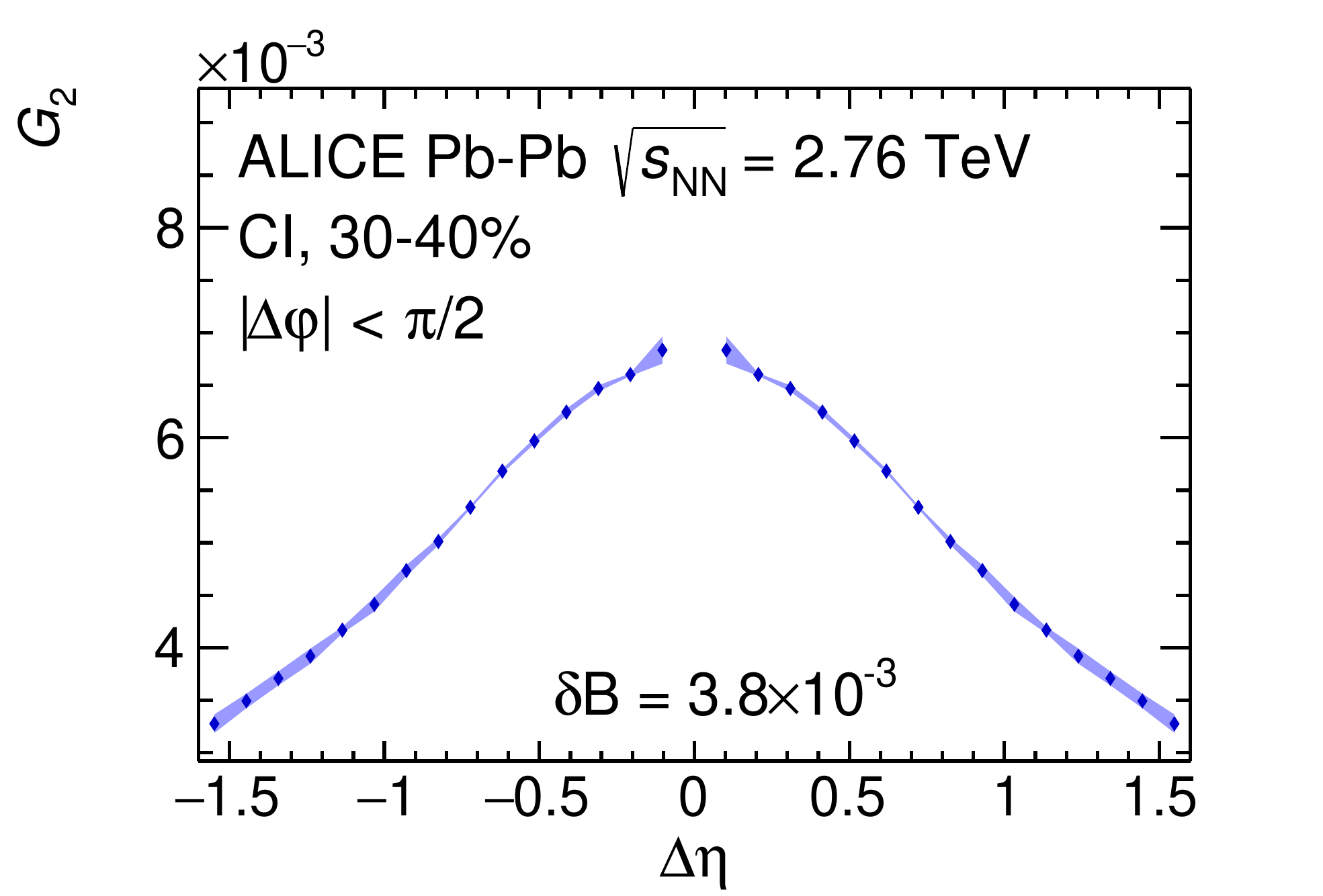}
  \includegraphics[scale=0.29,keepaspectratio=true,clip=true,trim=37pt 4pt 43pt 2pt]
  {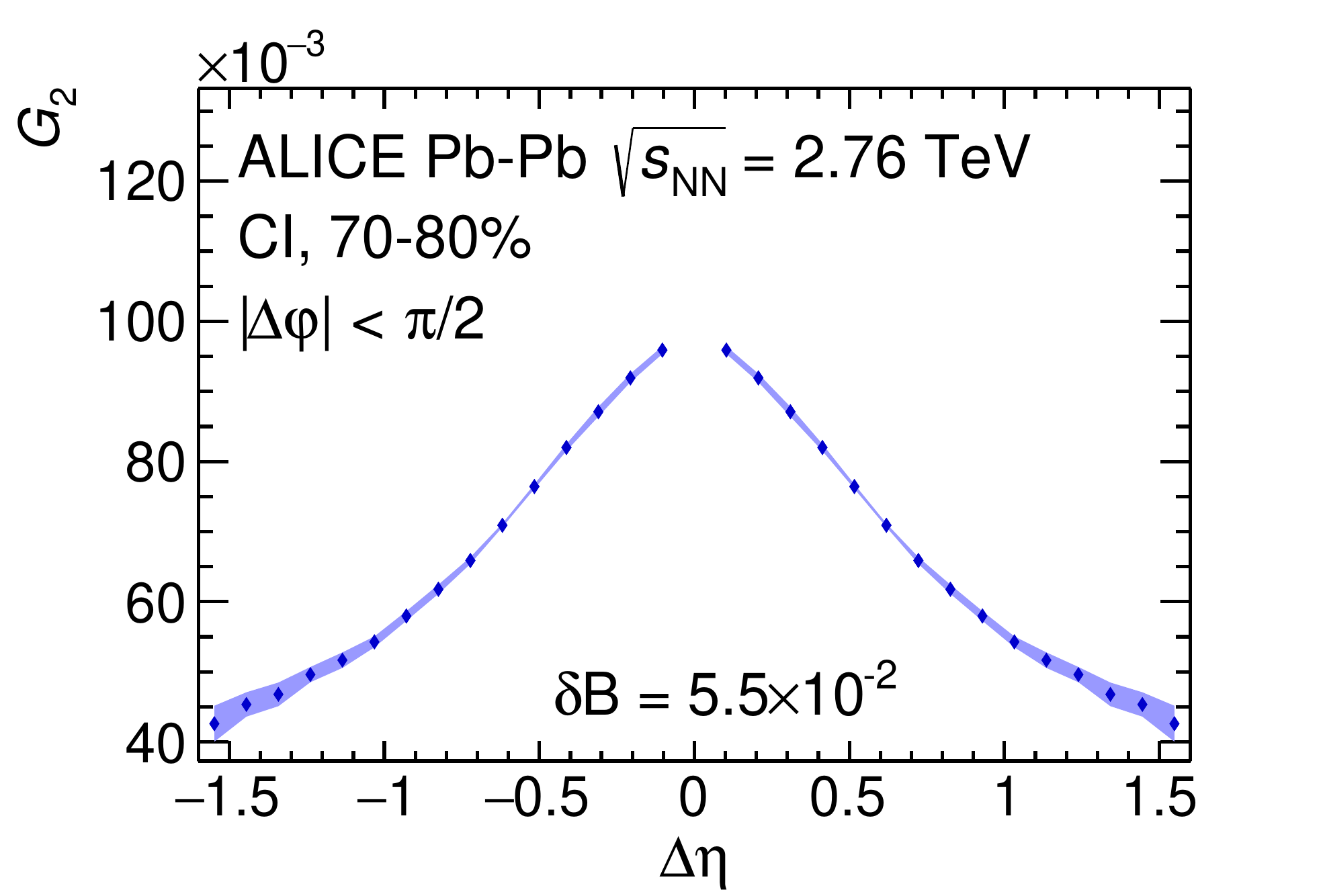} \par
  \includegraphics[scale=0.29,keepaspectratio=true,clip=true,trim=2pt 4pt 43pt 2pt]
  {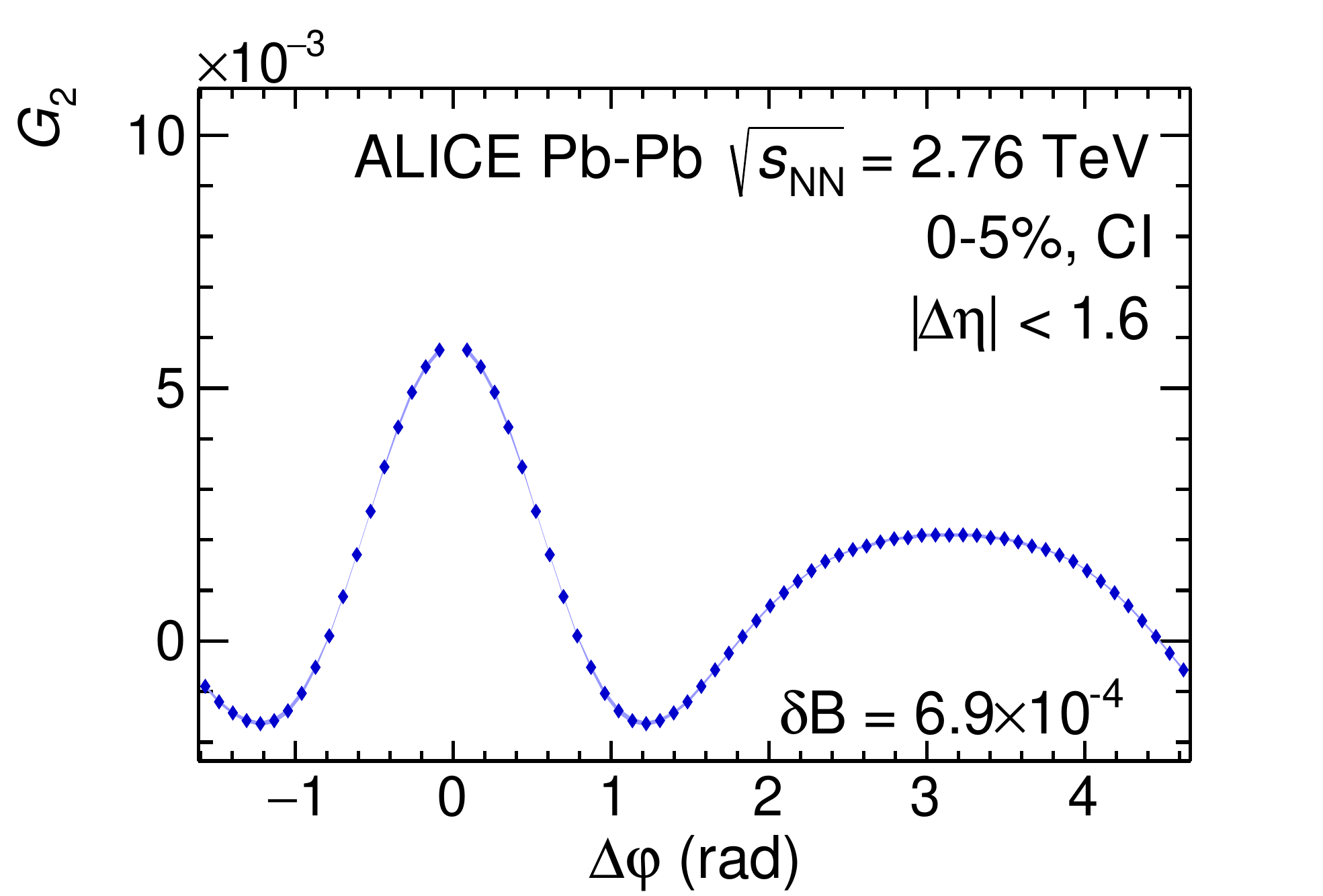}
  \includegraphics[scale=0.29,keepaspectratio=true,clip=true,trim=37pt 4pt 43pt 2pt]
  {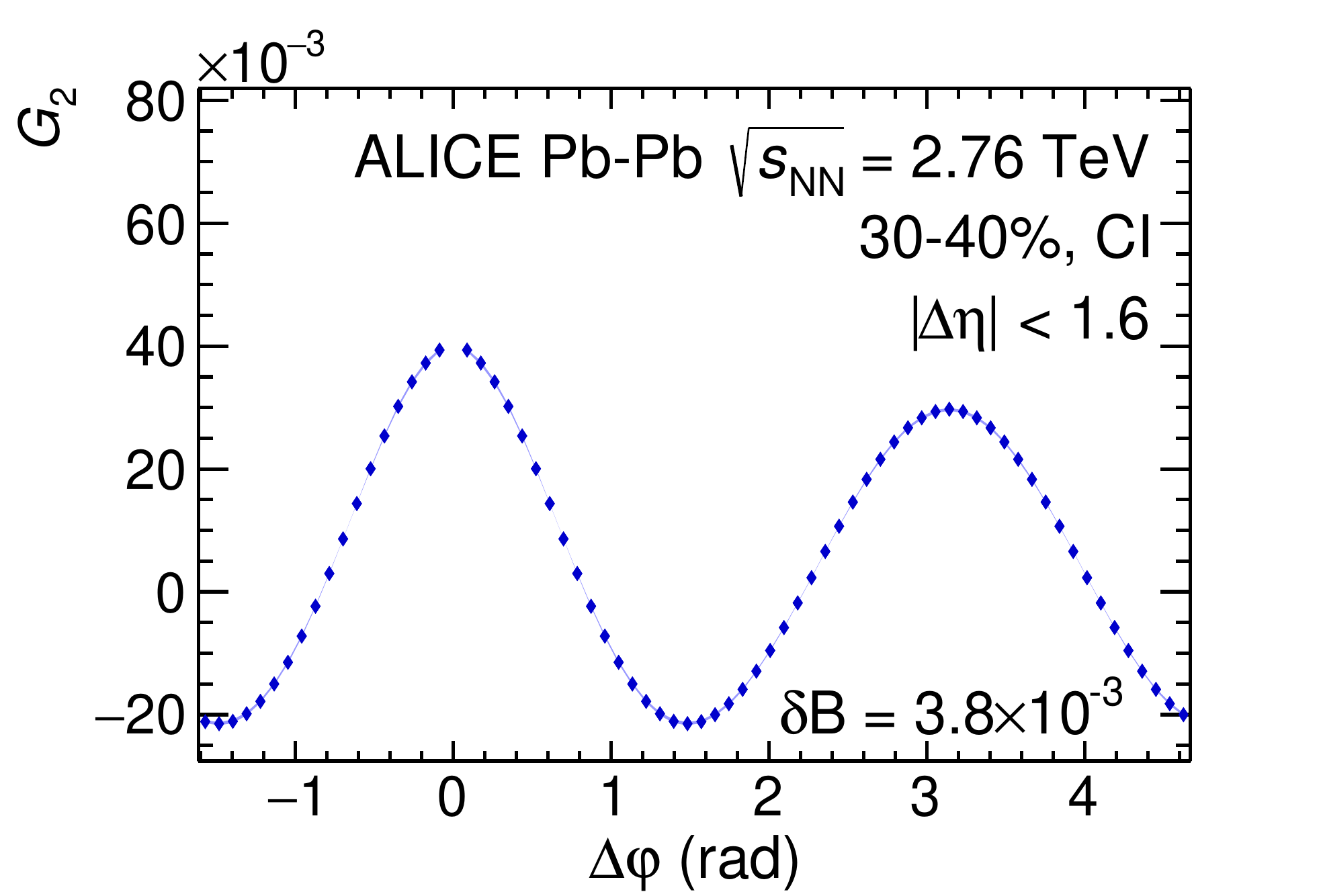}
  \includegraphics[scale=0.29,keepaspectratio=true,clip=true,trim=37pt 4pt 43pt 2pt]
  {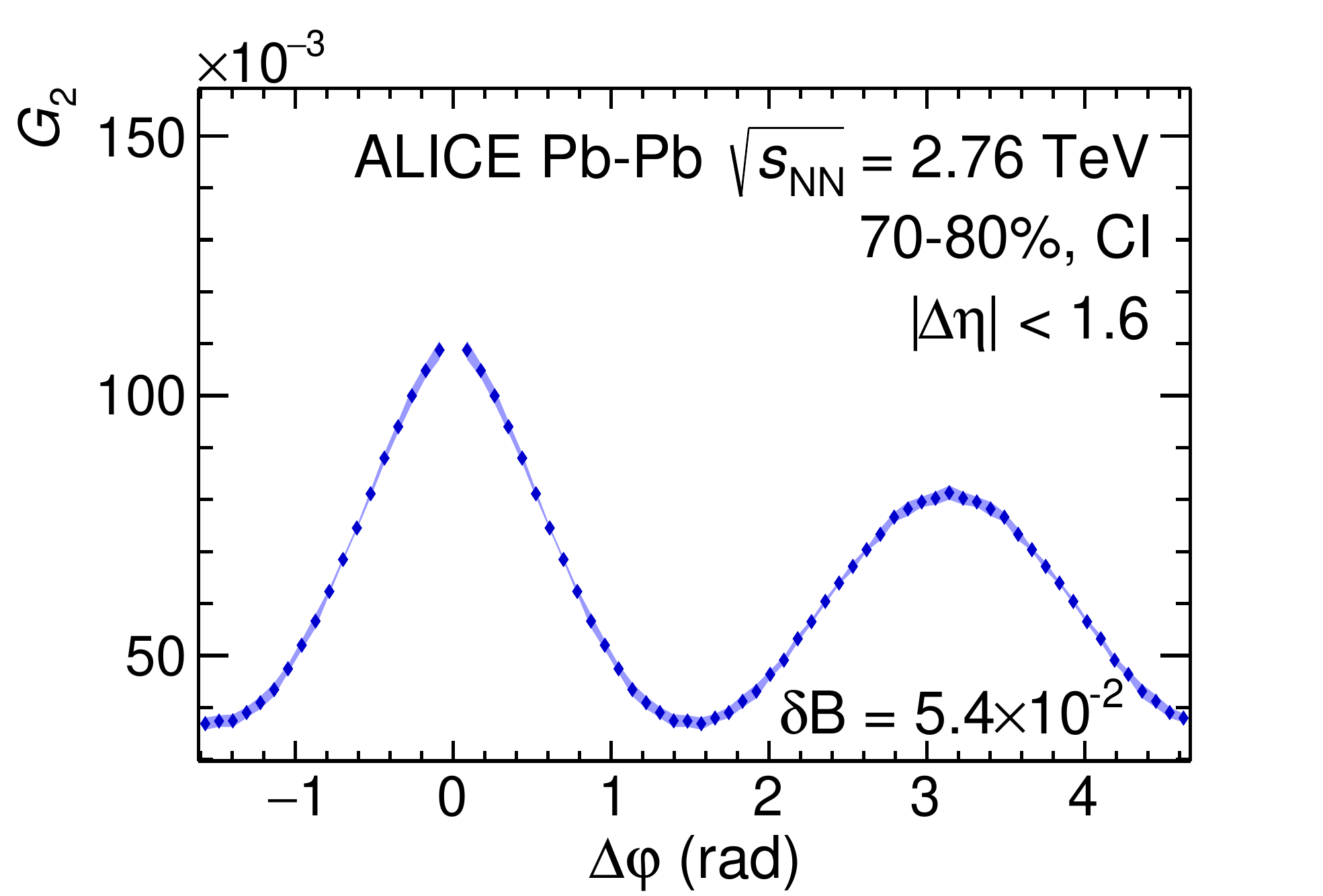}
  \caption{\label{fig:2dg2ci} Two-particle transverse momentum correlations $G_{2}^{\rm CI}$ (top) and their longitudinal  (middle) and azimuthal (bottom) projections for the most central (left), semi-central (center) and peripheral (right) \PbPb\ collisions at $\sqrt{s_{\rm NN}}=2.76\;\text{\TeVe}$. 
  Vertical bars (mostly smaller than the marker size) and shaded blue bands represent statistical and systematic uncertainties, respectively. The systematic uncertainty on the long-range mean correlator strength is quoted as $\delta B$ in both projections. Under-corrected correlator values at $\Delta\eta,\Delta\varphi=0$ are not shown. See text for details.}
\end{figure}

Figure~\ref{fig:2dg2ci} presents  the correlators   $G_2^{\rm CI}(\Delta \eta,\Delta \varphi)$ measured  in 0--5\%, 30--40\%, 70--80\%  Pb--Pb collisions, and their respective projections along the $\Delta\eta$ and $\Delta\varphi$ axes. 
The $G_2^{\rm CI}$ correlators feature sizable $\Delta \varphi$ modulations, dominated in mid-central collisions by a strong elliptic ($\cos (2\Delta\varphi)$) component. On the near-side, atop the azimuthal modulation,  the $G_2^{\rm CI}$ correlators  feature  a  near-side peak whose amplitude  monotonically decreases from peripheral to central collisions while its longitudinal width systematically broadens. Qualitatively similar trends were observed for the $R_2$ and $P_2$ correlators reported by ALICE~\cite{Acharya:2018ddg} and the $G_2^{\rm CI}$ correlator (there named $C$) reported by STAR~\cite{Agakishiev:2011fs}.  In most central collisions, the amplitude of the $\Delta\varphi$ modulations associated with collective flow decreases but the longitudinal broadening remains. Additionally, a depletion centered at $(\Delta \eta,\Delta \varphi) =(0, 0)$  consistent with previous ALICE results~\cite{Adam:2016tsv,Adam:2016ckp} can be seen.

\begin{figure*}[t]
  \centering
  \includegraphics[scale=0.80,keepaspectratio=true,clip=true,trim=0pt 0pt 15pt 10pt]
  {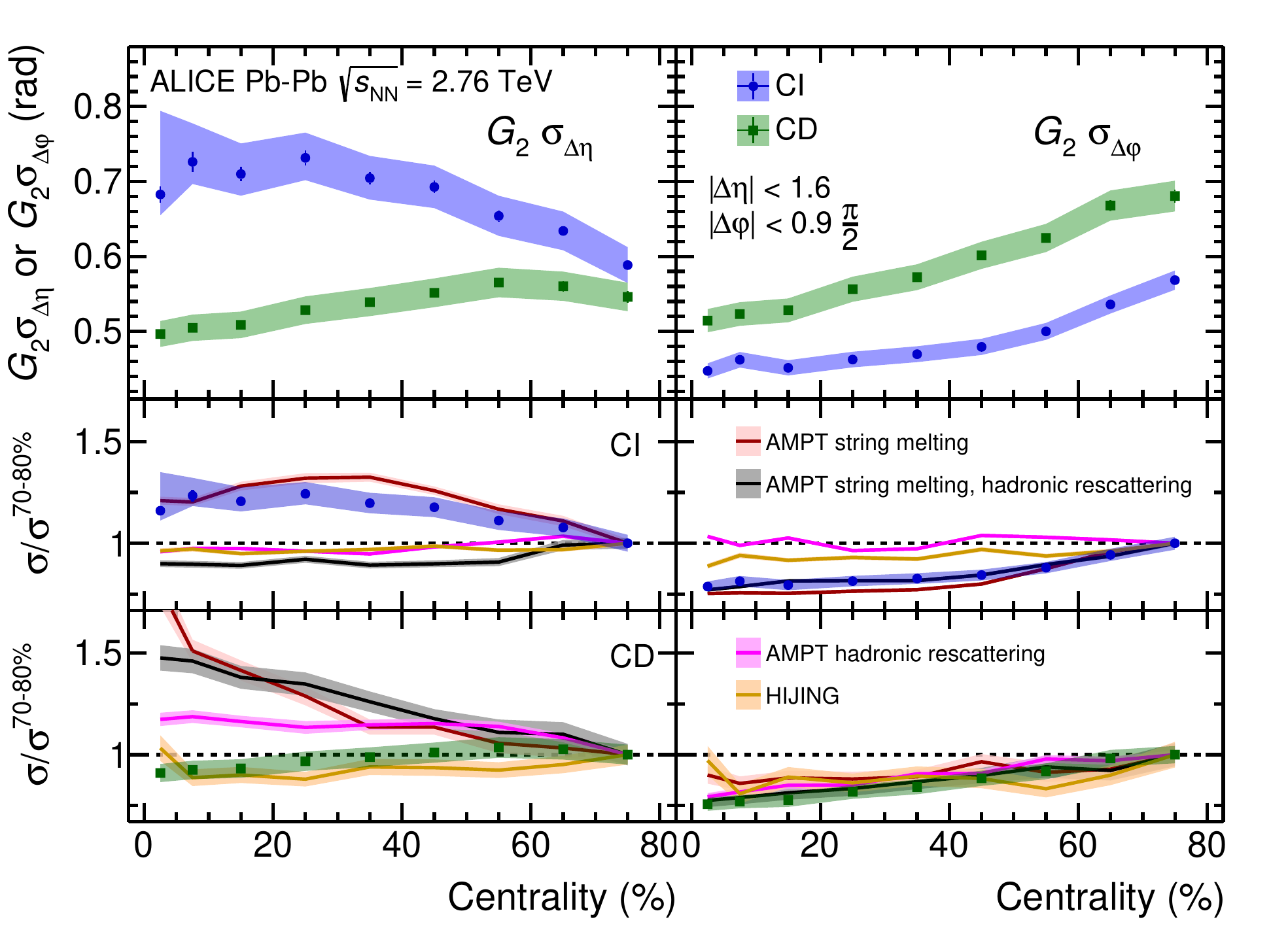}
  \caption{\label{fig:widthsevowithmodels} Top panels: collision centrality evolution of the longitudinal (left) and azimuthal  (right) widths of the $G_{2}$ CD and CI correlators measured in Pb--Pb collisions at $\sqrt{s_{\rm{NN}}} = 2.76\;\text{\TeVe}$. 
  Central and bottom panels: width evolution relative to the value in the most peripheral collisions of the two-particle transverse momentum correlations $G_{2}^{\rm CI}$ (central) and $G_{2}^{\rm CD}$ (bottom) along the longitudinal (left) and azimuthal (right) dimensions. Data are compared with HIJING and AMPT model expectations. In data, vertical bars and shaded bands represent statistical and systematic uncertainties, respectively. For models, shaded bands represent statistical uncertainties.}
\end{figure*}

In order to study the centrality evolution of the near-side peak of the $G_2^{\text CI}$ and $G_2^{\text CD}$ correlators independently of the underlying collective azimuthal behavior, they are separately parametrized with  a  two-component  model defined as
\begin{equation}
          F( \Delta \eta, \Delta \varphi) = B + \displaystyle\sum_{n=2}^{6} a_{n} \times \cos \left(n \Delta \varphi\right) + 
  A \times\frac{\gamma_{\Delta \eta}}{2\,\omega_{\Delta \eta}\,
      \Gamma \left(\frac{1}{\gamma_{\Delta \eta}}\right)}
      \,\rm{e}^{- \left| \frac{\Delta \eta}{\omega_{\Delta \eta}}\right|^{\gamma_{\Delta \eta}}}\times
          \frac{\gamma_{\Delta \varphi}}{2\,\omega_{\Delta \varphi}\,
      \Gamma \left(\frac{1}{\gamma_{\Delta \varphi}}\right)}
      \,\rm{e}^{- \left| \frac{\Delta \varphi}{\omega_{\Delta \varphi}}\right|^{\gamma_{\Delta \varphi}}} \text{,}
\label{eq:fitfunc}
\end{equation}
where $B$ and $a_n$ are intended to describe the  long-range mean correlation strength and azimuthal aniso\-tropy, while the bidimensional generalized Gaussian, defined by the parameters $A$, $\omega_{\Delta \eta}$, $\omega_{\Delta \varphi}$, $\gamma_{\Delta \eta}$ and $\gamma_{\Delta \varphi}$, is intended to model the signal of interest. The $(\Delta\eta,\Delta\varphi)=(0,0)$ depletion present in the $G_{2}^{\text{CI}}$ correlator is not properly modeled by Eq.~(\ref{eq:fitfunc}) and the depletion area, $|\Delta\eta|<0.31$ and $|\Delta\varphi|<0.26\;\text{rad.}$, is excluded from the fit.  Bidimensional fits are carried out considering only statistical uncertainties. In the case of  the $G_{2}^{\text{CI}}$ correlator the $\chi^2/\text{ndf}$ values for semi-central to peripheral collisions are found in the range 1--2; for central collisions they increase to 4. The area which contributes the most to the increase of the $\chi^2/\text{ndf}$  is the region between the generalized Gaussian and the Fourier expansion. Excluding this area the $\chi^2/\text{ndf}$ values obtained in central collisions are within the range 1--2.3. Fits of $G_{2}^{\text{CD}}$ give $\chi^2/\text{ndf}$ of the order of unity for peripheral to semi-central collisions and in the range 2--3.5 for  central collisions. Larger $\chi^2/\text{ndf}$ values observed in central collisions rise because the near side peak starts to depart from the generalized Gaussian description. 
The actual focus is on the evolution of the widths. The longitudinal and azimuthal widths of the correlators, denoted $\sigma_{\Delta \eta}$ and $\sigma_{\Delta \varphi}$, respectively, are  then extracted as the standard deviation of the generalized Gaussian 
\begin{equation}
  \sigma_{\Delta \eta (\Delta \varphi)} = \sqrt{\frac{\omega^2_{\Delta \eta (\Delta \varphi)} 
      \Gamma(3/\gamma_{\Delta \eta (\Delta \varphi)})}{\Gamma(1/\gamma_{\Delta \eta (\Delta \varphi)})}}\text{,}
\end{equation}
and plotted as a function of collision centrality in the top panels of Fig.~\ref{fig:widthsevowithmodels} for both $G_2^{\text{CI}}$ and $G_2^{\text{CD}}$ correlators.  
The global shift of the correlator strength, quoted as a systematic uncertainty in the projections of the correlators, does not affect the shape of the near-side peak of $G_2$. Accordingly, the widths are not affected either. 
Correlations between the contributors to the longitudinal width and the harmonic parameters for the $G_2^{\text{CI}}$ correlator are found as follows: $a_{2}$ and $a_{4}$ are anti-correlated with $\omega_{\Delta\eta}$ with values in the ranges $-$0.8 to $-$0.4 and $-$0.5--0, respectively, while $a_{3}$ is correlated with values 0--0.4. On the other hand, $a_{2}$ and $a_{4}$ are correlated with $\gamma_{\Delta\eta}$ with values within 0.4--0.8 and 0--0.5, respectively, while $a_{3}$ is anti-correlated with values in the range $-$0.5--0. $a_{2}$ correlations show no centrality dependence while the absolute value of $a_{3}$ and $a_{4}$ correlations decreases from central to peripheral collisions.
In the case of the contributors to the azimuthal width, $a_{2}$ and $a_{4}$ are correlated with $\omega_{\Delta\varphi}$ and with $\gamma_{\Delta\varphi}$ with values in the ranges 0.5--0.8 and 0.6--0.9, and 0.6--0.9 and 0.7--0.9, respectively, while $a_3$ is anti-correlated with both with values within $-$0.8 to $-$0.5 and $-$0.9 to $-$0.7. On the azimuthal dimension the absolute value of the harmonic coefficients correlations decreases towards peripheral collisions.
Systematic uncertainties in the widths of the near-side peak of $G_2^{\text{CI}}$ and $G_2^{\text{CD}}$ are mainly due to the presence of secondary particles. With the alternative track selection criterion, systematic uncertainties on the  longitudinal and azimuthal  widths of the near-side peak are estimated to be 2\% and 3\%, respectively, for both $G_2^{\text{CI}}$ and $G_2^{\text{CD}}$, for most central events, with decreasing values towards  peripheral collisions. Uncertainty contributions on the widths are not correlated with centrality and averages along centrality classes are considered.
Overall,  maximum systematic uncertainties of 4\%(2\%) and 3.5\%(3\%)  are assigned   to the $G_2^{\text{CI}}$ and $G_2^{\text{CD}}$ widths, respectively, along the longitudinal (azimuthal) dimension. The impact of the size of the area excluded from the fit on the  width of the $G_2^{\text{CI}}$ correlator is evaluated enlarging the area in both dimensions. Only semi-central to central centrality classes have their corresponding longitudinal widths modified. The effect is a broadening from 1.5\% in the 30--40\% class up to a broadening of 20\% in the 0--5\% class incorporated as an additional asymmetric systematic uncertainty on the widths of $G_2^{\text{CI}}$. On the azimuthal widths the impact is reduced to a 2\% narrowing.

\section{Discussion}
\begin{figure*}[ht]
  \includegraphics[scale=0.84,keepaspectratio=true,clip=true,trim=10pt 0pt 25pt 10pt]
  {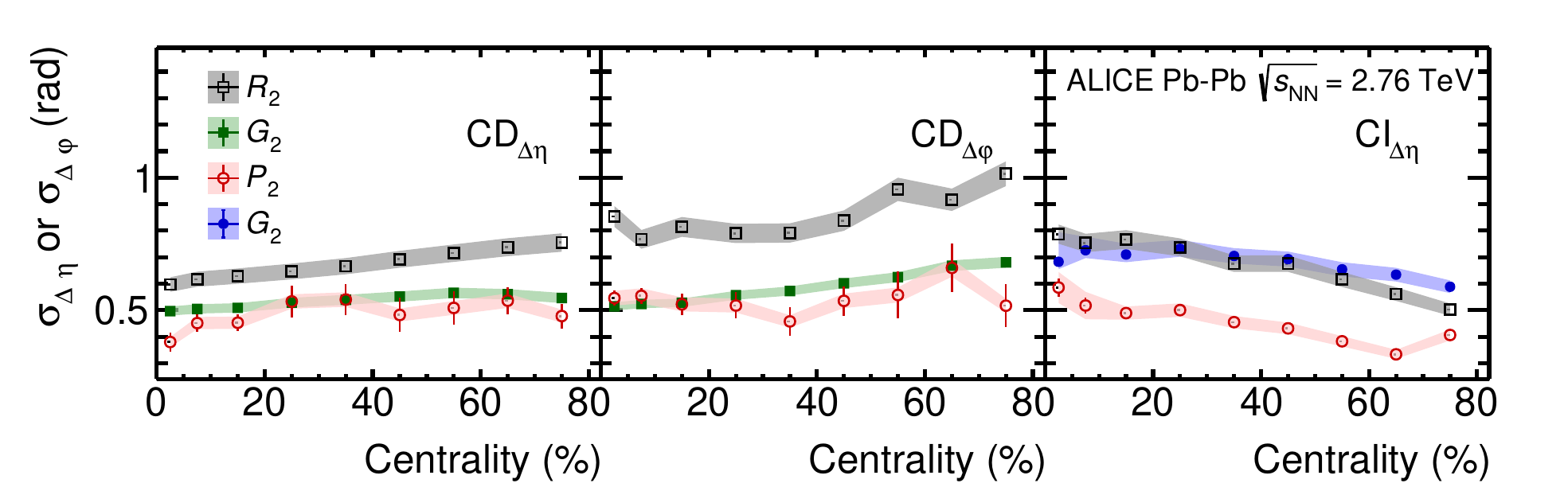}
  \caption{\label{fig:r2p2comp} Left panel: collision centrality evolution of the longitudinal width of  number correlator $R_2^{\rm CD}$ and transverse momentum correlators $P_2^{\rm CD}$ and $G_2^{\rm CD}$. Central panel: idem for the azimuthal width of $R_2^{\rm CD}$, $P_2^{\rm CD}$ and $G_2^{\rm CD}$. 
  Right panel: collision centrality evolution of the longitudinal width of $R_2^{\rm CI}$, $P_2^{\rm CI}$, and $G_2^{\rm CI}$. Data for $R_2$ and $P_2$ are from ~\cite{Acharya:2018ddg}.  Vertical bars and shaded  bands represent statistical and systematic uncertainties, respectively.}
\end{figure*}

Broadening and narrowing are hereafter intended as the behavior of the correlation function, measured by its widths, when going from peripheral collisions, high values of centrality percentile, to central collisions, lower values of centrality percentile.
The $G_{2}^{\rm{CI}}$ correlator broadens longitudinally but narrows in azimuth, whereas the $G_{2}^{\rm{CD}}$ correlator narrows both longitudinally and azimuthally.
As shown in Fig.~\ref{fig:r2p2comp}, these dependencies are qualitatively consistent with those of  $R_2$ and  $P_2$ correlators  measured  in the same kinematic range by the ALICE collaboration~\cite{Acharya:2018ddg}. Note that the $G_{2}$ correlator is sensitive to  transverse momentum and number density fluctuations since both affect the momentum current density. 
In contrast, $R_{2}$ is sensitive to number density fluctuations and $P_{2}$, sensitive to transverse momentum fluctuations, is designed to minimize the contribution of those number density fluctuations~\cite{Gavin:2008ta}. In fact~\cite{Gavin:2008ta} 
\begin{equation}
\left(P_{2}+1\right)\left(R_{2}+1\right) 
    = \left(G_{2} +1 \right)
\end{equation}
so, the increase in transverse momentum currents could be due to either the increase in multiplicity or the increase of transverse momentum.
The $G_{2}^{\rm{CD}}$ and $P_{2}^{\rm{CD}}$ correlators  feature approximately equal widths while $R_{2}^{\rm{CD}}$ is approximately 30\% wider throughout  its centrality evolution. The centrality dependence of  $G_{2}^{\rm{CD}}$ is qualitatively consistent with that of balance function (BF) observations~\cite{Abelev:2013csa,Adam:2015gda}. Phenomenological analyses of the BFs suggest that their narrowing with centrality is largely due to the presence of strong radial flow and  delayed hadronization  in \PbPb\ collisions~\cite{Abelev:2013csa}. 
It is thus reasonable to infer that radial flow and larger $\langle p_{\rm T}\rangle$, in more central collisions,  also produce the  observed narrowing of $G_{2}^{\rm{CD}}$. 
This conjecture is  supported by calculations of the collision centrality dependence of $G_{2}^{\rm{CD}}$ azimuthal widths  with the HIJING and AMPT models shown in the bottom right panel of Fig.~\ref{fig:widthsevowithmodels}. 
Radial flow might also explain the observed azimuthal narrowing of the $G_{2}^{\rm{CI}}$ correlator  with centrality, which is reasonably well reproduced by calculations with AMPT with string melting, but not by  HIJING or AMPT calculations with only hadronic rescattering as shown in central right panel of Fig.~\ref{fig:widthsevowithmodels}. 

The broadening of the longitudinal width of the $G_{2}^{\rm{CI}}$ correlator is of particular interest given predictions that it should grow in proportion to $\eta/s$ of the matter produced in the collisions~\cite{Gavin:2006xd}. 
As expected for a system with finite viscosity, it is found that $G_{2}^{\rm{CI}}$ broadens significantly with increasing collision centrality, while by contrast, $G_{2}^{\rm{CD}}$ exhibits a slight but distinct narrowing. 
This $G_{2}^{\rm{CD}}$ longitudinal narrowing is expected from a boost of particle pairs by radial flow but is not properly accounted for by AMPT calculations shown in the bottom left panel of Fig.~\ref{fig:widthsevowithmodels}. 
Radial flow should also produce a narrowing of the $G_{2}^{\rm{CI}}$ correlator in the longitudinal direction. However competing effects, possibly associated with the finite shear viscosity of the system, are instead producing a significant broadening although reaching what seems a saturation level at semi-central collisions. Note that HIJING and AMPT, with the hadronic rescattering enabled, grossly fail to reproduce the observed broadening and instead predict a slight narrowing (Fig.~\ref{fig:widthsevowithmodels} central left panel). AMPT with string melting and without the hadronic rescattering phase qualitatively reproduces the longitudinal broadening of $G_{2}^{\rm{CI}}$, even its saturation, but grossly miss the narrowing of $G_{2}^{\rm{CD}}$ along that dimension and thus cannot be considered reliable in this context.  

Particles produced by jet fragmentation are also known to exhibit correlations and jet-medium interactions can broaden such correlations.
Two-particle correlation measurements, of particles associated with high-$p_{\rm T}$ jets, indeed show substantial broadening of low $p_{\rm T}$ particle correlations  relative to correlation functions measured in pp collisions~\cite{Chatrchyan:2014ava,Adam:2016ckp,Adam:2016tsv}. 
This broadening, however, is observed  in both the longitudinal and azimuthal directions in stark contrast with the behavior of the inclusive $G_{2}^{\rm{CI}}$  correlator measured in this work which exhibits a significant narrowing in the azimuthal direction. Additionally,  the number of particles from jets is relatively small compared to the number from  the bulk.
Therefore, although jet fragmentation may contribute to the broadening observed in the longitudinal direction, it is unlikely to amount to a significant contribution given the observed narrowing in the $\Delta\varphi$ direction and the relatively low impact of correlations from jet particles.  

\begin{figure}[ht]
  \centering
  \includegraphics[scale=0.50,keepaspectratio=true,clip=true,trim=2pt 4pt 50pt 2pt]{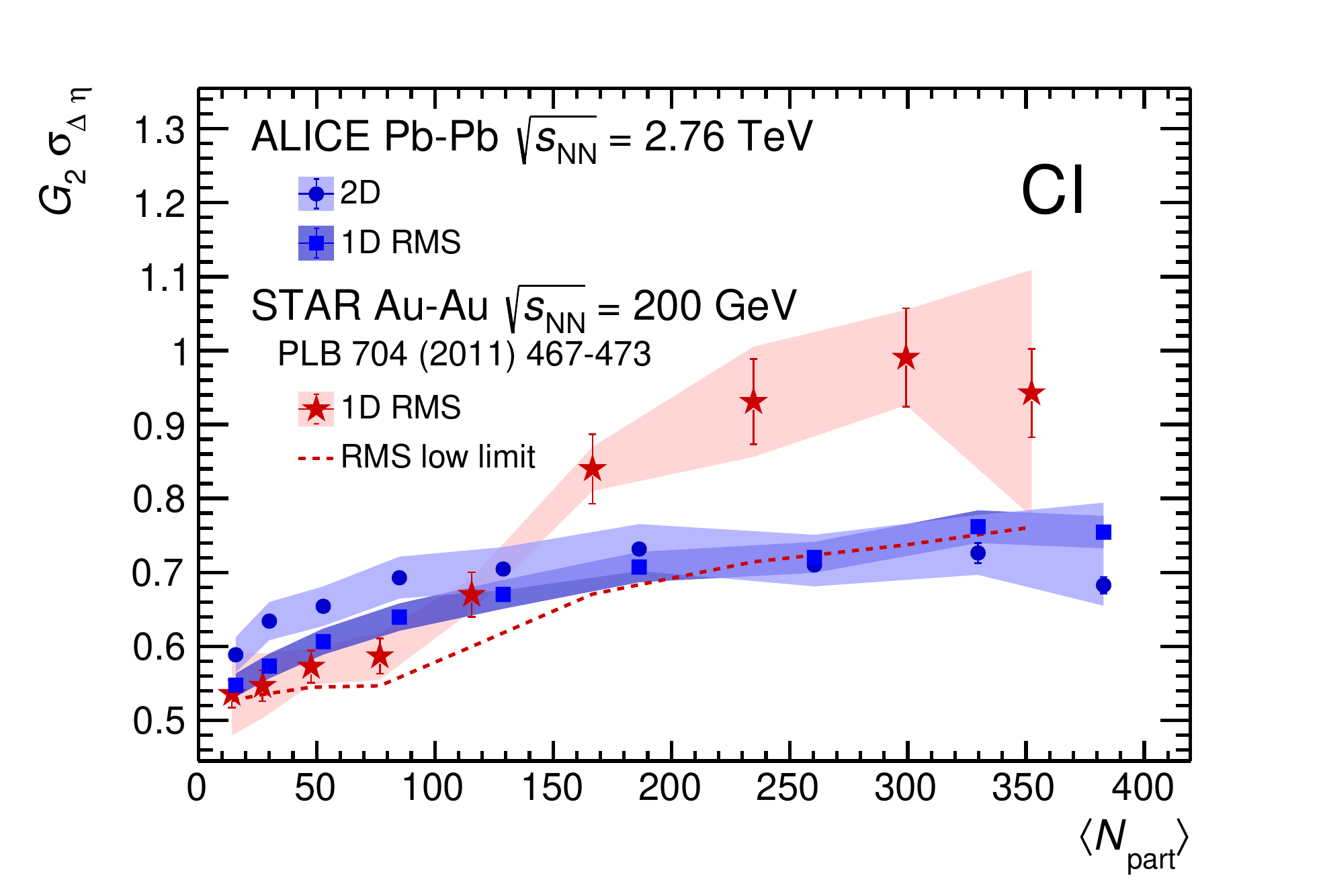}
  \caption{\label{fig:starcomp} Two-particle transverse  momentum correlation $G_{2}^{\rm{CI}}$ longitudinal width evolution with the
  number of participants in Au--Au collisions at $\sqrt{s_{\rm{NN}}}=200\;\text{\GeVe}$~\cite{Agakishiev:2011fs}  and in Pb--Pb collisions at $\sqrt{s_{\rm{NN}}}=2.76\;\text{\TeVe}$, measured in this work,  using  the bi-dimensional fit described in the text (2D) and the  method used by the STAR experiment~\cite{Agakishiev:2011fs}~(1D). For completeness, STAR RMS low limit~\cite{Agakishiev:2011fs} is also shown.}
\end{figure}

Figure~\ref{fig:starcomp} compares  results from this analysis with those reported by the STAR collaboration~\cite{Agakishiev:2011fs}. For proper comparison,  Fig.~\ref{fig:starcomp} presents root mean square (RMS) widths  of  $\Delta\eta$  projections of $G_2^{\rm CI}$ calculated  above a long range baseline  as in the STAR analysis~\cite{Agakishiev:2011fs}.
Although STAR reported results are based on the dimensional version of $G_{2}^{\rm CI}$, the same expression as in Eq.~(\ref{eq:normG2}) but without the normalization $\langle p_{{\rm T},1} \rangle \langle p_{{\rm T},2} \rangle$, the correlator widths reported in this letter are identical for both, the dimensional and dimensionless versions of the $G_{2}$ correlator. The longitudinal broadening measured in this analysis, using the 1D RMS method, amounts to 36\% while that observed by STAR reaches 74\% showing also a saturation at semi-central collisions. 
It was verified that the smaller broadening seen in this analysis is not a result of the slightly  narrower longitudinal acceptance of the ALICE experiment by testing the analysis method with Monte Carlo models reproducing the approximate shape and strength of the measured correlation functions. The longitudinal broadening of $G_2^{\rm CI}$ and its observed saturation thus appears to be potentially dependent on the beam energy.

\begin{figure}[ht]
  \centering
  \includegraphics[scale=0.50,keepaspectratio=true,clip=true,trim=2pt 4pt 50pt 10pt]
  {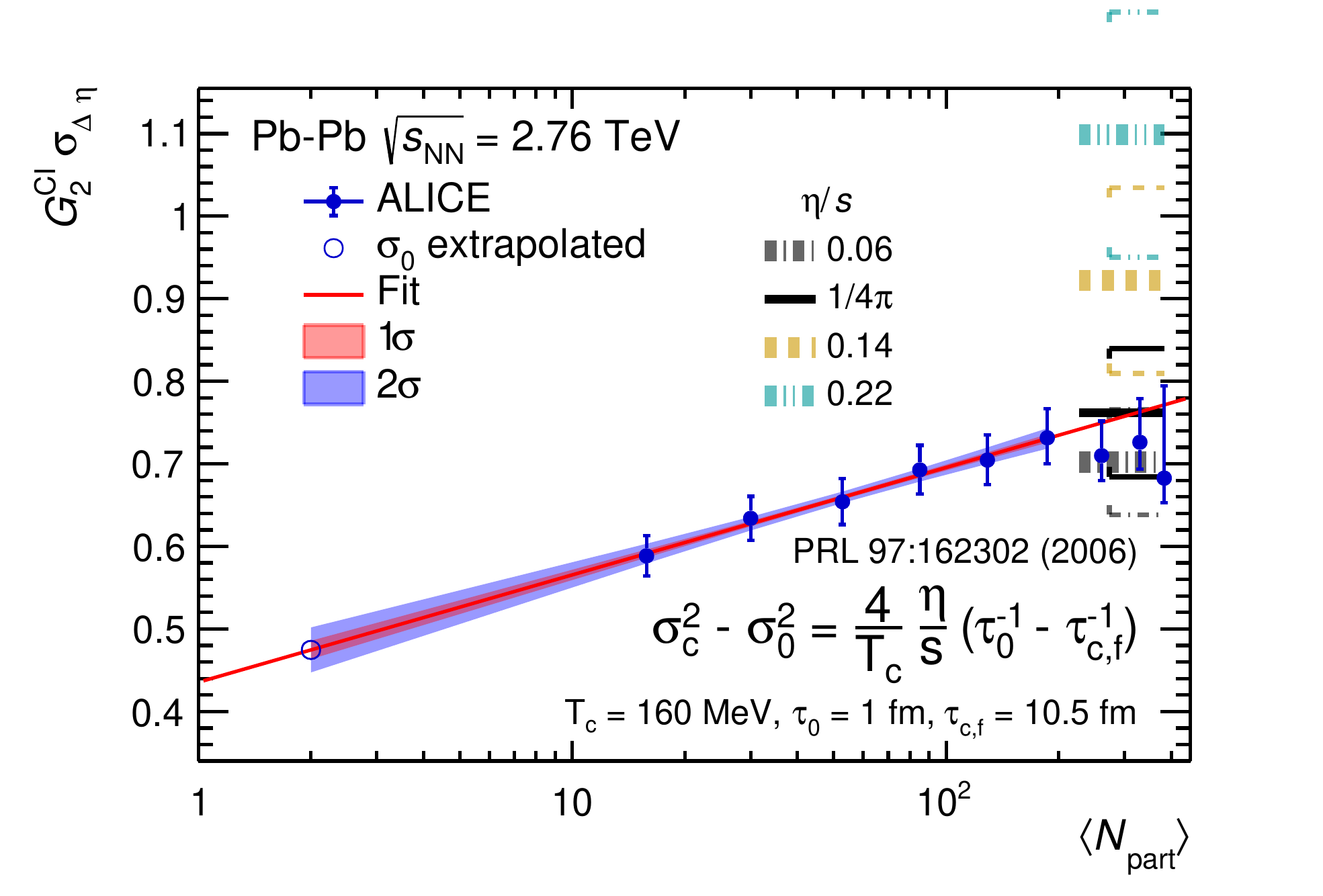}
  \caption{ Expected longitudinal widths for the most central collisions of the two-particle transverse momentum correlation $G_{2}^{\rm{CI}}$ for different values of $\eta/s$ by using the expression suggested in~\cite{Gavin:2006xd}. Data point error bars represent total uncertainties obtained by adding in quadrature statistical and systematic uncertainties. In the formula $\sigma_{\rm c}$ is the longitudinal width for the most central collisions inferred by using this expression and represented for each of the $\etas$ values by the color discontinuous bands (continuous for $\eta/s=1/4\pi$) at the highest number of participants, $\sigma_{0}$ is the longitudinal width for the most peripheral collisions (only two participants) which is obtained by extrapolating the fit, $T_{\rm c}$ is the critical temperature, $\tau_{0}$ is the formation time and $\tau_{\rm c,f}$ the freeze-out time. Error caps in the same color as the discontinuous bands, represent uncertainties of the inferred longitudinal widths for the most central collisions (see text for details).}
  \label{fig:shearviscosity}
\end{figure}

Interpreting the longitudinal broadening of $G_{2}^{\rm CI}$ as originating exclusively from viscous effects, an estimate of the shear viscosity per unit of entropy density, $\eta/s$,  of the matter produced in heavy-ion collisions can be extracted~\cite{Agakishiev:2011fs} using the expression 
\begin{equation}
    \sigma_{\rm c}^{2} - \sigma_{0}^{2} = \frac{4}{T_{\rm c}}
        \,\frac{\eta}{s}\,\left(\frac{1}{\tau_{0}} - \frac{1}{\tau_{\rm c,f}}\right)
  \label{eq:sigmacentral}
\end{equation}
derived in~\cite{Gavin:2006xd}.  
In Eq.~\eqref{eq:sigmacentral} $\sigma_{\rm c}$ is the longitudinal width for the most central collisions (ideally 0\% centrality), $\sigma_{0}$ is the longitudinal width for the most peripheral collisions (ideally 100\% centrality), $T_{\rm c}$ is the critical temperature, $\tau_{0}$ is the formation time and $\tau_{\rm c,f}$ the freeze-out time. 
The correlator width for the most peripheral \PbPb collisions at $\sqrt{s_{\rm NN}} = 2.76\;\text{TeV}$ is estimated based on a power law extrapolation of the measured values, shown in Fig.~\ref{fig:shearviscosity}, down to  $N_{\rm part}=2$. 
Canonical values are used for the critical temperature, $T_{\rm c}=160$ \MeVe~\cite{Becattini:2014rea},  the formation time $\tau_0=1\;\text{fm}/c$~\cite{Becattini:2014rea}, and the freeze-out time,  $\tau_{\rm c,f} =10.5$ fm/$c$~\cite{Aamodt:2011mr}.  
With these inputs in Eq.~\eqref{eq:sigmacentral}, $G_2^{\rm CI}$ longitudinal widths for the most central collisions are calculated for several values of $\eta/s=0.06$, $1/4\pi$, $0.14$ and $0.22$ and also shown in Fig.~\ref{fig:shearviscosity} as color discontinuous (continuous for $\eta/s=1/4\pi$) bands at the highest number of participants. 
Considering 2\%, 30\%, and 3\% uncertainties for $T_{\rm c}$ ($155 < T_{\rm c} < 165\;\text{TeV}$), $\tau_{0}$, and $\tau_{\rm c,f}$ ($10 < \tau_{\rm c,f} < 11\;\text{fm}$) respectively, the uncertainties of the four obtained $G_{2}^{\rm CI}$ longitudinal widths for the most central collisions reach 9\%, 10\%, 12\%, and 14\%, respectively, also shown in Fig.~\ref{fig:shearviscosity} as error caps in the same color as the discontinuous bands.
The   $G_2^{\rm CI}$ correlator width measured in central collisions thus favors rather small  values of $\eta/s$, close to the KSS limit of $1/4\pi$~\cite{Kovtun:2004de}.
The authors of Ref.~\cite{Gavin:2006xd} obtain the correlator width values, for Au--Au collisions at $\sqrt{s_{\rm NN}} = 200\;\text{GeV}$, without an actual measurement of $G_{2}^{\rm CI}$ from the only available two-particle transverse momentum correlator which in its turn was inferred from event-wise mean transverse momentum fluctuations~\cite{Adams:2005aw} and on its energy dependence~\cite{Adams:2006sg}. They constrain $\eta/s$ to a relatively wide interval 0.08--0.30. The precision of the STAR measurement is limited by the relative uncertainty of the $G_{2}^{\rm CI}$ correlator widths for Au--Au collisions at $\sqrt{s_{\rm NN}} = 200\;\text{GeV}$; $\eta/s = \text{0.06--0.21}$ was reported in~\cite{Agakishiev:2011fs}. 

\section{Conclusions}
Measurements of charge dependent (CD) and charge independent (CI) transverse momentum correlators $G_2$ in \PbPb\ collisions at $\sqrt{s_{\rm NN}} = 2.76$ \TeVe were presented aiming at the determination of the shear viscosity per unit of entropy density, $\eta/s$, of the matter formed in such collisions. 
The near-side peak of the $G_2^{\rm CD}$ correlator is observed to significantly narrow with collision centrality both in the longitudinal and azimuthal directions. This behavior is found to be similar to that of the charge balance function  as a result, most likely, of an increase of the average radial flow velocity from peripheral to central collisions. 
By contrast, the $G_2^{\rm CI}$ correlator is found to narrow only in the azimuthal direction with collision centrality and features a sizable broadening in the longitudinal direction. The observed broadening along the longitudinal direction is expected based on friction forces associated with the finite shear viscosity of the system. Taking the model proposed in~\cite{Gavin:2006xd}, an estimate of the value of $\eta/s$ of  order $1/4\pi$, in qualitative agreement with values obtained from other methods~\cite{Bernhard:2016tnd,Moreland:2018gsh}, is obtained. 
String melting AMPT without the hadronic rescattering phase has been found to qualitatively reproduce the longitudinal broadening of $G_{2}^{\rm{CI}}$ but grossly misses the narrowing of $G_{2}^{\rm{CD}}$ along that dimension.  
The observed saturation in the longitudinal broadening and the sizable difference in  broadening  relative to that observed by STAR may result from the interplay of viscous forces and kinematic narrowing associated to radial flow. 
In the latter case, the difference compared to the STAR results due to a possible dependence on the beam energy could be better established with expanded experimental measurements for energies in the beam energy scan (BES) at RHIC or at 5.02 \TeVe at the LHC.

\newenvironment{acknowledgement}{\relax}{\relax}
\begin{acknowledgement}
\section*{Acknowledgements}
Authors thank Dr. Sean Gavin and Dr. George Moschelli for fruitful discussions.

The ALICE Collaboration would like to thank all its engineers and technicians for their invaluable contributions to the construction of the experiment and the CERN accelerator teams for the outstanding performance of the LHC complex.
The ALICE Collaboration gratefully acknowledges the resources and support provided by all Grid centres and the Worldwide LHC Computing Grid (WLCG) collaboration.
The ALICE Collaboration acknowledges the following funding agencies for their support in building and running the ALICE detector:
A. I. Alikhanyan National Science Laboratory (Yerevan Physics Institute) Foundation (ANSL), State Committee of Science and World Federation of Scientists (WFS), Armenia;
Austrian Academy of Sciences, Austrian Science Fund (FWF): [M 2467-N36] and Nationalstiftung f\"{u}r Forschung, Technologie und Entwicklung, Austria;
Ministry of Communications and High Technologies, National Nuclear Research Center, Azerbaijan;
Conselho Nacional de Desenvolvimento Cient\'{\i}fico e Tecnol\'{o}gico (CNPq), Financiadora de Estudos e Projetos (Finep), Funda\c{c}\~{a}o de Amparo \`{a} Pesquisa do Estado de S\~{a}o Paulo (FAPESP) and Universidade Federal do Rio Grande do Sul (UFRGS), Brazil;
Ministry of Education of China (MOEC) , Ministry of Science \& Technology of China (MSTC) and National Natural Science Foundation of China (NSFC), China;
Ministry of Science and Education and Croatian Science Foundation, Croatia;
Centro de Aplicaciones Tecnol\'{o}gicas y Desarrollo Nuclear (CEADEN), Cubaenerg\'{\i}a, Cuba;
Ministry of Education, Youth and Sports of the Czech Republic, Czech Republic;
The Danish Council for Independent Research | Natural Sciences, the VILLUM FONDEN and Danish National Research Foundation (DNRF), Denmark;
Helsinki Institute of Physics (HIP), Finland;
Commissariat \`{a} l'Energie Atomique (CEA), Institut National de Physique Nucl\'{e}aire et de Physique des Particules (IN2P3) and Centre National de la Recherche Scientifique (CNRS) and R\'{e}gion des  Pays de la Loire, France;
Bundesministerium f\"{u}r Bildung und Forschung (BMBF) and GSI Helmholtzzentrum f\"{u}r Schwerionenforschung GmbH, Germany;
General Secretariat for Research and Technology, Ministry of Education, Research and Religions, Greece;
National Research, Development and Innovation Office, Hungary;
Department of Atomic Energy Government of India (DAE), Department of Science and Technology, Government of India (DST), University Grants Commission, Government of India (UGC) and Council of Scientific and Industrial Research (CSIR), India;
Indonesian Institute of Science, Indonesia;
Centro Fermi - Museo Storico della Fisica e Centro Studi e Ricerche Enrico Fermi and Istituto Nazionale di Fisica Nucleare (INFN), Italy;
Institute for Innovative Science and Technology , Nagasaki Institute of Applied Science (IIST), Japanese Ministry of Education, Culture, Sports, Science and Technology (MEXT) and Japan Society for the Promotion of Science (JSPS) KAKENHI, Japan;
Consejo Nacional de Ciencia (CONACYT) y Tecnolog\'{i}a, through Fondo de Cooperaci\'{o}n Internacional en Ciencia y Tecnolog\'{i}a (FONCICYT) and Direcci\'{o}n General de Asuntos del Personal Academico (DGAPA), Mexico;
Nederlandse Organisatie voor Wetenschappelijk Onderzoek (NWO), Netherlands;
The Research Council of Norway, Norway;
Commission on Science and Technology for Sustainable Development in the South (COMSATS), Pakistan;
Pontificia Universidad Cat\'{o}lica del Per\'{u}, Peru;
Ministry of Science and Higher Education and National Science Centre, Poland;
Korea Institute of Science and Technology Information and National Research Foundation of Korea (NRF), Republic of Korea;
Ministry of Education and Scientific Research, Institute of Atomic Physics and Ministry of Research and Innovation and Institute of Atomic Physics, Romania;
Joint Institute for Nuclear Research (JINR), Ministry of Education and Science of the Russian Federation, National Research Centre Kurchatov Institute, Russian Science Foundation and Russian Foundation for Basic Research, Russia;
Ministry of Education, Science, Research and Sport of the Slovak Republic, Slovakia;
National Research Foundation of South Africa, South Africa;
Swedish Research Council (VR) and Knut \& Alice Wallenberg Foundation (KAW), Sweden;
European Organization for Nuclear Research, Switzerland;
Suranaree University of Technology (SUT), National Science and Technology Development Agency (NSDTA) and Office of the Higher Education Commission under NRU project of Thailand, Thailand;
Turkish Atomic Energy Agency (TAEK), Turkey;
National Academy of  Sciences of Ukraine, Ukraine;
Science and Technology Facilities Council (STFC), United Kingdom;
National Science Foundation of the United States of America (NSF) and United States Department of Energy, Office of Nuclear Physics (DOE NP), United States of America.    
\end{acknowledgement}
\bibliographystyle{utphys}   
\bibliography{bibliography}

\providecommand{\noopsort}[1]{}\providecommand{\singleletter}[1]{#1}%
\providecommand{\href}[2]{#2}\begingroup\raggedright\begin{thebibliography}{10}

\bibitem{Adams:2005dq}
{\bfseries STAR} Collaboration, J.~Adams {\em et~al.}, ``{Experimental and
  theoretical challenges in the search for the quark gluon plasma: The STAR
  Collaboration's critical assessment of the evidence from RHIC collisions}'',
  \href{http://dx.doi.org/10.1016/j.nuclphysa.2005.03.085}{{\em Nucl. Phys.}
  {\bfseries A757} (2005) 102--183},
  \href{http://arxiv.org/abs/nucl-ex/0501009}{{\ttfamily arXiv:nucl-ex/0501009
  [nucl-ex]}}.

\bibitem{Adcox:2004mh}
{\bfseries PHENIX} Collaboration, K.~Adcox {\em et~al.}, ``{Formation of dense
  partonic matter in relativistic nucleus-nucleus collisions at RHIC:
  Experimental evaluation by the PHENIX Collaboration}'',
  \href{http://dx.doi.org/10.1016/j.nuclphysa.2005.03.086}{{\em Nucl. Phys.}
  {\bfseries A757} (2005) 184--283},
\href{http://arxiv.org/abs/nucl-ex/0410003}{{\ttfamily arXiv:nucl-ex/0410003
  [nucl-ex]}}.

\bibitem{Arsene20051}
{\bfseries BRAHMS} Collaboration, I.~Arsene {\em et~al.}, ``{Quark gluon plasma
  and color glass condensate at RHIC? The perspective from the BRAHMS
  experiment}'',
  \href{http://dx.doi.org/http://dx.doi.org/10.1016/j.nuclphysa.2005.02.130}{{\em
  Nuclear Physics A} {\bfseries 757} no.~12, (2005) 1 -- 27}.
  \url{http://www.sciencedirect.com/science/article/pii/S0375947405002770}.
  First Three Years of Operation of RHIC.

\bibitem{Back:2004je}
{\bfseries PHOBOS} Collaboration, B.~B. Back {\em et~al.}, ``{The PHOBOS
  perspective on discoveries at RHIC}'',
  \href{http://dx.doi.org/10.1016/j.nuclphysa.2005.03.084}{{\em Nucl. Phys.}
  {\bfseries A757} (2005) 28--101},
  \href{http://arxiv.org/abs/nucl-ex/0410022}{{\ttfamily arXiv:nucl-ex/0410022
  [nucl-ex]}}.

\bibitem{Heinz:2011kt}
U.~Heinz, C.~Shen, and H.~Song, ``{The viscosity of quark-gluon plasma at RHIC
  and the LHC}'', \href{http://dx.doi.org/10.1063/1.3700674}{{\em AIP Conf.
  Proc.} {\bfseries 1441} no.~1, (2012) 766--770},
\href{http://arxiv.org/abs/1108.5323}{{\ttfamily arXiv:1108.5323 [nucl-th]}}.

\bibitem{Adams:2004bi}
{\bfseries STAR} Collaboration, J.~Adams {\em et~al.}, ``{Azimuthal anisotropy
  in Au+Au collisions at $\sqrt{s_{\rm NN}} = 200\;\text{\GeVe}$}'',
  \href{http://dx.doi.org/10.1103/PhysRevC.72.014904}{{\em Phys. Rev.}
  {\bfseries C72} (2005) 014904},
\href{http://arxiv.org/abs/nucl-ex/0409033}{{\ttfamily arXiv:nucl-ex/0409033
  [nucl-ex]}}.

\bibitem{Aamodt:2011by}
{\bfseries ALICE} Collaboration, K.~Aamodt {\em et~al.}, ``{Harmonic
  decomposition of two-particle angular correlations in Pb--Pb collisions at
  $\sqrt{s_{\rm NN}}=$ 2.76 TeV}'',
  \href{http://dx.doi.org/10.1016/j.physletb.2012.01.060}{{\em Phys.\ Lett.}
  {\bfseries B708} (2012) 249--264},
  \href{http://arxiv.org/abs/1109.2501}{{\ttfamily arXiv:1109.2501 [nucl-ex]}}.

\bibitem{Heinz:2013th}
U.~Heinz and R.~Snellings, ``{Collective flow and viscosity in relativistic
  heavy-ion collisions}'',
  \href{http://dx.doi.org/10.1146/annurev-nucl-102212-170540}{{\em Ann. Rev.
  Nucl. Part. Sci.} {\bfseries 63} (2013) 123--151},
\href{http://arxiv.org/abs/1301.2826}{{\ttfamily arXiv:1301.2826 [nucl-th]}}.

\bibitem{PhysRevLett.105.252302}
{\bfseries ALICE} Collaboration, K.~Aamodt {\em et~al.}, ``{Elliptic flow of
  charged particles in Pb--Pb collisions at 2.76 TeV}'',
  \href{http://dx.doi.org/10.1103/PhysRevLett.105.252302}{{\em Phys. Rev.
  Lett.} {\bfseries 105} (2010) 252302},
\href{http://arxiv.org/abs/1011.3914}{{\ttfamily arXiv:1011.3914 [nucl-ex]}}.

\bibitem{Song:2010mg}
H.~Song, S.~A. Bass, U.~Heinz, T.~Hirano, and C.~Shen, ``{200 A GeV Au+Au
  collisions serve a nearly perfect quark-gluon liquid}'',
  \href{http://dx.doi.org/10.1103/PhysRevLett.106.192301,
  10.1103/PhysRevLett.109.139904}{{\em Phys. Rev. Lett.} {\bfseries 106} (2011)
  192301}, \href{http://arxiv.org/abs/1011.2783}{{\ttfamily arXiv:1011.2783
  [nucl-th]}}.
[Erratum: Phys. Rev. Lett. 109, 139904 (2012)].

\bibitem{Shen:2012vn}
C.~Shen and U.~Heinz, ``{Collision Energy Dependence of Viscous Hydrodynamic
  Flow in Relativistic Heavy-Ion Collisions}'',
  \href{http://dx.doi.org/10.1103/PhysRevC.86.049903,
  10.1103/PhysRevC.85.054902}{{\em Phys. Rev.} {\bfseries C85} (2012) 054902},
  \href{http://arxiv.org/abs/1202.6620}{{\ttfamily arXiv:1202.6620 [nucl-th]}}.
[Erratum: Phys. Rev.C86,049903(2012)].

\bibitem{Acharya:2017gsw}
{\bfseries ALICE} Collaboration, S.~Acharya {\em et~al.}, ``{Systematic studies
  of correlations between different order flow harmonics in Pb--Pb collisions
  at $\sqrt{s_{\rm NN}}$ = 2.76 TeV}'',
  \href{http://dx.doi.org/10.1103/PhysRevC.97.024906}{{\em Phys. Rev.}
  {\bfseries C97} no.~2, (2018) 024906},
\href{http://arxiv.org/abs/1709.01127}{{\ttfamily arXiv:1709.01127 [nucl-ex]}}.

\bibitem{PhysRevC.97.044905}
J.~Auvinen, J.~E. Bernhard, S.~A. Bass, and I.~Karpenko, ``Investigating the
  collision energy dependence of $\ensuremath{\eta}/s$ in the beam energy scan
  at the {BNL Relativistic Heavy Ion Collider using Bayesian} statistics'',
  \href{http://dx.doi.org/10.1103/PhysRevC.97.044905}{{\em Phys. Rev. C}
  {\bfseries 97} (Apr, 2018) 044905}.
  \url{https://link.aps.org/doi/10.1103/PhysRevC.97.044905}.

\bibitem{Bernhard:2016tnd}
J.~E. Bernhard, J.~S. Moreland, S.~A. Bass, J.~Liu, and U.~Heinz, ``{{Applying
  Bayesian parameter estimation to relativistic heavy-ion collisions:
  simultaneous characterization of the initial state and quark-gluon plasma
  medium}}'', \href{http://dx.doi.org/10.1103/PhysRevC.94.024907}{{\em Phys.
  Rev.} {\bfseries C94} no.~2, (2016) 024907},
\href{http://arxiv.org/abs/1605.03954}{{\ttfamily arXiv:1605.03954 [nucl-th]}}.

\bibitem{Gavin:2006xd}
S.~Gavin and M.~Abdel-Aziz, ``{Measuring Shear Viscosity Using Transverse
  Momentum Correlations in Relativistic Nuclear Collisions}'',
  \href{http://dx.doi.org/10.1103/PhysRevLett.97.162302}{{\em Phys. Rev. Lett.}
  {\bfseries 97} (2006) 162302},
\href{http://arxiv.org/abs/nucl-th/0606061}{{\ttfamily arXiv:nucl-th/0606061
  [nucl-th]}}.

\bibitem{Agakishiev:2011fs}
{\bfseries STAR} Collaboration, G.~Agakishiev {\em et~al.}, ``{Evolution of the
  differential transverse momentum correlation function with centrality in
  Au$+$Au collisions at $\sqrt{s_{\rm NN}} =$ 200 GeV}'',
  \href{http://dx.doi.org/10.1016/j.physletb.2011.09.075}{{\em Phys. Lett.}
  {\bfseries B704} (2011) 467--473},
\href{http://arxiv.org/abs/1106.4334}{{\ttfamily arXiv:1106.4334 [nucl-ex]}}.

\bibitem{Ravan:2013lwa}
S.~Ravan, P.~Pujahari, S.~Prasad, and C.~A. Pruneau, ``{Correcting Correlation
  Function Measurements}'',
  \href{http://dx.doi.org/10.1103/PhysRevC.89.024906}{{\em Phys. Rev.}
  {\bfseries C89} no.~2, (2014) 024906},
\href{http://arxiv.org/abs/1311.3915}{{\ttfamily arXiv:1311.3915 [nucl-ex]}}.

\bibitem{Acharya:2018ddg}
{\bfseries ALICE} Collaboration, S.~Acharya {\em et~al.}, ``{Two-particle
  differential transverse momentum and number density correlations in p--Pb and
  Pb--Pb at the LHC}'',
  \href{http://dx.doi.org/10.1103/PhysRevC.100.044903}{{\em Phys. Rev. C}
  {\bfseries 100} (Oct, 2019) 044903},
\href{http://arxiv.org/abs/1805.04422}{{\ttfamily arXiv:1805.04422 [nucl-ex]}}.

\bibitem{Gonzalez:2018cty}
V.~Gonzalez, A.~Marin, P.~Ladron De~Guevara, J.~Pan, S.~Basu, and C.~Pruneau,
  ``{Effect of centrality bin width corrections on two-particle number and
  transverse momentum differential correlation functions}'',
  \href{http://dx.doi.org/10.1103/PhysRevC.99.034907}{{\em Phys. Rev.}
  {\bfseries C99} no.~3, (2019) 034907},
\href{http://arxiv.org/abs/1809.04962}{{\ttfamily arXiv:1809.04962
  [physics.data-an]}}.

\bibitem{Sharma:2008qr}
M.~Sharma and C.~A. Pruneau, ``{Methods for the Study of Transverse Momentum
  Differential Correlations}'',
  \href{http://dx.doi.org/10.1103/PhysRevC.79.024905}{{\em Phys. Rev.}
  {\bfseries C79} (2009) 024905},
\href{http://arxiv.org/abs/0810.0716}{{\ttfamily arXiv:0810.0716 [nucl-ex]}}.

\bibitem{1748-0221-3-08-S08002}
{\bfseries ALICE} Collaboration, K.~Aamodt {\em et~al.}, ``{The ALICE
  experiment at the CERN LHC}'', {\em Journal of Instrumentation} {\bfseries 3}
  no.~08, (2008) S08002. \url{http://stacks.iop.org/1748-0221/3/i=08/a=S08002}.

\bibitem{Abelev:2014ffa}
{\bfseries ALICE} Collaboration, B.~Abelev {\em et~al.}, ``{Performance of the
  ALICE Experiment at the CERN LHC}'',
  \href{http://dx.doi.org/10.1142/S0217751X14300440}{{\em Int.J.Mod.Phys.}
  {\bfseries A29} (2014) 1430044},
\href{http://arxiv.org/abs/1402.4476}{{\ttfamily arXiv:1402.4476 [nucl-ex]}}.

\bibitem{Abelev:2013qoq}
{\bfseries ALICE} Collaboration, B.~Abelev {\em et~al.}, ``{Centrality
  determination of Pb--Pb collisions at $\sqrt{s_{\rm NN}}$ = 2.76 TeV with
  ALICE}'', \href{http://dx.doi.org/10.1103/PhysRevC.88.044909}{{\em Phys.
  Rev.} {\bfseries C88} no.~4, (2013) 044909},
\href{http://arxiv.org/abs/1301.4361}{{\ttfamily arXiv:1301.4361 [nucl-ex]}}.

\bibitem{Wang:1991hta}
X.-N. Wang and M.~Gyulassy, ``{HIJING}: A {M}onte {C}arlo model for multiple
  jet production in p p, p {A} and {A} {A} collisions'',
\href{http://dx.doi.org/10.1103/PhysRevD.44.3501}{{\em Phys. Rev.} {\bfseries
  D44} (1991) 3501--3516}.

\bibitem{Brun:1082634}
R.~Brun, F.~Bruyant, F.~Carminati, S.~Giani, M.~Maire, A.~McPherson,
  G.~Patrick, and L.~Urban,
  \href{http://dx.doi.org/10.17181/CERN.MUHF.DMJ1}{{\em {GEANT: Detector
  Description and Simulation Tool; Oct 1994}}}.
\newblock CERN Program Library. CERN, Geneva, 1993.
\newblock \url{http://cds.cern.ch/record/1082634}.
\newblock Long Writeup W5013.

\bibitem{Barlow:2002yb}
R.~Barlow, ``{Systematic errors: Facts and fictions}'', in {\em {Advanced
  Statistical Techniques in Particle Physics. Proceedings, Conference, Durham,
  UK, March 18-22, 2002}}, pp.~134--144.
\newblock 2002.
\newblock \href{http://arxiv.org/abs/hep-ex/0207026}{{\ttfamily
  arXiv:hep-ex/0207026 [hep-ex]}}.
\newblock
\url{http://www.ippp.dur.ac.uk/Workshops/02/statistics/proceedings//barlow.pdf}.
\newblock

\bibitem{Adam:2016tsv}
{\bfseries ALICE} Collaboration, J.~Adam {\em et~al.}, ``{Anomalous evolution
  of the near-side jet peak shape in Pb--Pb collisions at $\sqrt{s_{\rm NN}}$ =
  2.76 TeV}'', \href{http://dx.doi.org/10.1103/PhysRevLett.119.102301}{{\em
  Phys. Rev. Lett.} {\bfseries 119} no.~10, (2017) 102301},
\href{http://arxiv.org/abs/1609.06643}{{\ttfamily arXiv:1609.06643 [nucl-ex]}}.

\bibitem{Adam:2016ckp}
{\bfseries ALICE} Collaboration, J.~Adam {\em et~al.}, ``{Evolution of the
  longitudinal and azimuthal structure of the near-side jet peak in Pb--Pb
  collisions at $\sqrt{s_{\rm NN}} = 2.76$ TeV}'',
  \href{http://dx.doi.org/10.1103/PhysRevC.96.034904}{{\em Phys. Rev.}
  {\bfseries C96} no.~3, (2017) 034904},
\href{http://arxiv.org/abs/1609.06667}{{\ttfamily arXiv:1609.06667 [nucl-ex]}}.

\bibitem{Gavin:2008ta}
S.~Gavin and G.~Moschelli, ``{Viscosity and the Soft Ridge at RHIC}'',
  \href{http://dx.doi.org/10.1088/0954-3899/35/10/104084}{{\em J. Phys.}
  {\bfseries G35} (2008) 104084},
\href{http://arxiv.org/abs/0806.4366}{{\ttfamily arXiv:0806.4366 [nucl-th]}}.

\bibitem{Abelev:2013csa}
{\bfseries ALICE} Collaboration, B.~Abelev {\em et~al.}, ``{Charge correlations
  using the balance function in Pb--Pb collisions at $\sqrt{s_{\rm NN}}$ = 2.76
  TeV}'', \href{http://dx.doi.org/10.1016/j.physletb.2013.05.039}{{\em Phys.
  Lett.} {\bfseries B723} (2013) 267--279},
\href{http://arxiv.org/abs/1301.3756}{{\ttfamily arXiv:1301.3756 [nucl-ex]}}.

\bibitem{Adam:2015gda}
{\bfseries ALICE} Collaboration, J.~Adam {\em et~al.}, ``{Multiplicity and
  transverse momentum evolution of charge-dependent correlations in pp, p–Pb,
  and Pb–Pb collisions at the LHC}'',
  \href{http://dx.doi.org/10.1140/epjc/s10052-016-3915-1}{{\em Eur. Phys. J.}
  {\bfseries C76} no.~2, (2016) 86},
\href{http://arxiv.org/abs/1509.07255}{{\ttfamily arXiv:1509.07255 [nucl-ex]}}.

\bibitem{Chatrchyan:2014ava}
{\bfseries CMS} Collaboration, S.~Chatrchyan {\em et~al.}, ``{Measurement of
  jet fragmentation in PbPb and pp collisions at $\sqrt{s_{\rm NN}}=2.76$
  TeV}'', \href{http://dx.doi.org/10.1103/PhysRevC.90.024908}{{\em Phys. Rev.}
  {\bfseries C90} no.~2, (2014) 024908},
\href{http://arxiv.org/abs/1406.0932}{{\ttfamily arXiv:1406.0932 [nucl-ex]}}.

\bibitem{Becattini:2014rea}
F.~Becattini, ``{The Quark Gluon Plasma and relativistic heavy ion collisions
  in the LHC era}'',
\href{http://dx.doi.org/10.1088/1742-6596/527/1/012012}{{\em J. Phys. Conf.
  Ser.} {\bfseries 527} (2014) 012012}.

\bibitem{Aamodt:2011mr}
{\bfseries ALICE} Collaboration, K.~Aamodt {\em et~al.}, ``{Two-pion
  Bose-Einstein correlations in central Pb--Pb collisions at $\sqrt{s_{\rm
  NN}}$ = 2.76 \TeVe}'',
  \href{http://dx.doi.org/10.1016/j.physletb.2010.12.053}{{\em Phys. Lett.}
  {\bfseries B696} (2011) 328--337},
\href{http://arxiv.org/abs/1012.4035}{{\ttfamily arXiv:1012.4035 [nucl-ex]}}.

\bibitem{Kovtun:2004de}
P.~Kovtun, D.~T. Son, and A.~O. Starinets, ``{Viscosity in strongly interacting
  quantum field theories from black hole physics}'',
  \href{http://dx.doi.org/10.1103/PhysRevLett.94.111601}{{\em Phys. Rev. Lett.}
  {\bfseries 94} (2005) 111601},
\href{http://arxiv.org/abs/hep-th/0405231}{{\ttfamily arXiv:hep-th/0405231
  [hep-th]}}.

\bibitem{Adams:2005aw}
{\bfseries STAR} Collaboration, J.~Adams {\em et~al.}, ``{Transverse-momentum
  {$p_{\rm T}$} correlations on $(\eta, \varphi)$ from mean-{$p_{\rm T}$}
  fluctuations in {Au--Au} collisions at $\sqrt{s_{\rm NN}} =
  200\;\text{GeV}$}'', \href{http://dx.doi.org/10.1088/0954-3899/32/6/L02}{{\em
  J. Phys.} {\bfseries G32} (2006) L37--L48},
\href{http://arxiv.org/abs/nucl-ex/0509030}{{\ttfamily arXiv:nucl-ex/0509030
  [nucl-ex]}}.

\bibitem{Adams:2006sg}
{\bfseries STAR} Collaboration, J.~Adams {\em et~al.}, ``{The Energy dependence
  of $p_{\rm T}$ angular correlations inferred from mean-$p_{\rm T}$
  fluctuation scale dependence in heavy ion collisions at the SPS and RHIC}'',
  \href{http://dx.doi.org/10.1088/0954-3899/34/3/004}{{\em J. Phys.} {\bfseries
  G34} (2007) 451--466},
\href{http://arxiv.org/abs/nucl-ex/0605021}{{\ttfamily arXiv:nucl-ex/0605021
  [nucl-ex]}}.

\bibitem{Moreland:2018gsh}
J.~S. Moreland, J.~E. Bernhard, and S.~A. Bass, ``{Estimating initial state and
  quark-gluon plasma medium properties using a hybrid model with nucleon
  substructure calibrated to $p$-Pb and Pb--Pb collisions at
  $\sqrt{s_\mathrm{NN}}=5.02$ TeV}'',
\href{http://arxiv.org/abs/1808.02106}{{\ttfamily arXiv:1808.02106 [nucl-th]}}.

\end{thebibliography}\endgroup


\providecommand{\noopsort}[1]{}\providecommand{\singleletter}[1]{#1}%
\providecommand{\href}[2]{#2}\begingroup\raggedright\endgroup
\newpage
\appendix
\section{The ALICE Collaboration}
\label{app:collab}

\begingroup
\small
\begin{flushleft}
S.~Acharya\Irefn{org141}\And 
D.~Adamov\'{a}\Irefn{org94}\And 
A.~Adler\Irefn{org74}\And 
J.~Adolfsson\Irefn{org80}\And 
M.M.~Aggarwal\Irefn{org99}\And 
G.~Aglieri Rinella\Irefn{org33}\And 
M.~Agnello\Irefn{org30}\And 
N.~Agrawal\Irefn{org10}\textsuperscript{,}\Irefn{org53}\And 
Z.~Ahammed\Irefn{org141}\And 
S.~Ahmad\Irefn{org16}
% ...
\And
\renewcommand\labelenumi{\textsuperscript{\theenumi}~}

\section*{Affiliation notes}
\renewcommand\theenumi{\roman{enumi}}
\begin{Authlist}
\item \Adef{org*}Deceased
\item \Adef{orgI}Dipartimento DET del Politecnico di Torino, Turin, Italy
\item \Adef{orgII}M.V. Lomonosov Moscow State University, D.V. Skobeltsyn Institute of Nuclear, Physics, Moscow, Russia
\item \Adef{orgIII}Department of Applied Physics, Aligarh Muslim University, Aligarh, India
\item \Adef{orgIV}Institute of Theoretical Physics, University of Wroclaw, Poland
\end{Authlist}

\section*{Collaboration Institutes}
\renewcommand\theenumi{\arabic{enumi}~}
\begin{Authlist}
\item \Idef{org1}A.I. Alikhanyan National Science Laboratory (Yerevan Physics Institute) Foundation, Yerevan, Armenia
\item \Idef{org2}Bogolyubov Institute for Theoretical Physics, National Academy of Sciences of Ukraine, Kiev, Ukraine
\item \Idef{org3}Bose Institute, Department of Physics  and Centre for Astroparticle Physics and Space Science (CAPSS), Kolkata, India
\item \Idef{org4}Budker Institute for Nuclear Physics, Novosibirsk, Russia
\item \Idef{org5}California Polytechnic State University, San Luis Obispo, California, United States
\item \Idef{org6}Central China Normal University, Wuhan, China
\item \Idef{org7}Centre de Calcul de l'IN2P3, Villeurbanne, Lyon, France
\item \Idef{org8}Centro de Aplicaciones Tecnol\'{o}gicas y Desarrollo Nuclear (CEADEN), Havana, Cuba
\item \Idef{org9}Centro de Investigaci\'{o}n y de Estudios Avanzados (CINVESTAV), Mexico City and M\'{e}rida, Mexico
\item \Idef{org10}Centro Fermi - Museo Storico della Fisica e Centro Studi e Ricerche ``Enrico Fermi', Rome, Italy
\item \Idef{org11}Chicago State University, Chicago, Illinois, United States
\item \Idef{org12}China Institute of Atomic Energy, Beijing, China
\item \Idef{org13}Comenius University Bratislava, Faculty of Mathematics, Physics and Informatics, Bratislava, Slovakia
\item \Idef{org14}COMSATS University Islamabad, Islamabad, Pakistan
\item \Idef{org15}Creighton University, Omaha, Nebraska, United States
\item \Idef{org16}Department of Physics, Aligarh Muslim University, Aligarh, India
\item \Idef{org17}Department of Physics, Pusan National University, Pusan, Republic of Korea
\item \Idef{org18}Department of Physics, Sejong University, Seoul, Republic of Korea
\item \Idef{org19}Department of Physics, University of California, Berkeley, California, United States
\item \Idef{org20}Department of Physics, University of Oslo, Oslo, Norway
\item \Idef{org21}Department of Physics and Technology, University of Bergen, Bergen, Norway
\item \Idef{org22}Dipartimento di Fisica dell'Universit\`{a} 'La Sapienza' and Sezione INFN, Rome, Italy
\item \Idef{org23}Dipartimento di Fisica dell'Universit\`{a} and Sezione INFN, Cagliari, Italy
\item \Idef{org24}Dipartimento di Fisica dell'Universit\`{a} and Sezione INFN, Trieste, Italy
\item \Idef{org25}Dipartimento di Fisica dell'Universit\`{a} and Sezione INFN, Turin, Italy
\item \Idef{org26}Dipartimento di Fisica e Astronomia dell'Universit\`{a} and Sezione INFN, Bologna, Italy
\item \Idef{org27}Dipartimento di Fisica e Astronomia dell'Universit\`{a} and Sezione INFN, Catania, Italy
\item \Idef{org28}Dipartimento di Fisica e Astronomia dell'Universit\`{a} and Sezione INFN, Padova, Italy
\item \Idef{org29}Dipartimento di Fisica `E.R.~Caianiello' dell'Universit\`{a} and Gruppo Collegato INFN, Salerno, Italy
\item \Idef{org30}Dipartimento DISAT del Politecnico and Sezione INFN, Turin, Italy
\item \Idef{org31}Dipartimento di Scienze e Innovazione Tecnologica dell'Universit\`{a} del Piemonte Orientale and INFN Sezione di Torino, Alessandria, Italy
\item \Idef{org32}Dipartimento Interateneo di Fisica `M.~Merlin' and Sezione INFN, Bari, Italy
\item \Idef{org33}European Organization for Nuclear Research (CERN), Geneva, Switzerland
\item \Idef{org34}Faculty of Electrical Engineering, Mechanical Engineering and Naval Architecture, University of Split, Split, Croatia
\item \Idef{org35}Faculty of Engineering and Science, Western Norway University of Applied Sciences, Bergen, Norway
\item \Idef{org36}Faculty of Nuclear Sciences and Physical Engineering, Czech Technical University in Prague, Prague, Czech Republic
\item \Idef{org37}Faculty of Science, P.J.~\v{S}af\'{a}rik University, Ko\v{s}ice, Slovakia
\item \Idef{org38}Frankfurt Institute for Advanced Studies, Johann Wolfgang Goethe-Universit\"{a}t Frankfurt, Frankfurt, Germany
\item \Idef{org39}Fudan University, Shanghai, China
\item \Idef{org40}Gangneung-Wonju National University, Gangneung, Republic of Korea
\item \Idef{org41}Gauhati University, Department of Physics, Guwahati, India
\item \Idef{org42}Helmholtz-Institut f\"{u}r Strahlen- und Kernphysik, Rheinische Friedrich-Wilhelms-Universit\"{a}t Bonn, Bonn, Germany
\item \Idef{org43}Helsinki Institute of Physics (HIP), Helsinki, Finland
\item \Idef{org44}High Energy Physics Group,  Universidad Aut\'{o}noma de Puebla, Puebla, Mexico
\item \Idef{org45}Hiroshima University, Hiroshima, Japan
\item \Idef{org46}Hochschule Worms, Zentrum  f\"{u}r Technologietransfer und Telekommunikation (ZTT), Worms, Germany
\item \Idef{org47}Horia Hulubei National Institute of Physics and Nuclear Engineering, Bucharest, Romania
\item \Idef{org48}Indian Institute of Technology Bombay (IIT), Mumbai, India
\item \Idef{org49}Indian Institute of Technology Indore, Indore, India
\item \Idef{org50}Indonesian Institute of Sciences, Jakarta, Indonesia
\item \Idef{org51}INFN, Laboratori Nazionali di Frascati, Frascati, Italy
\item \Idef{org52}INFN, Sezione di Bari, Bari, Italy
\item \Idef{org53}INFN, Sezione di Bologna, Bologna, Italy
\item \Idef{org54}INFN, Sezione di Cagliari, Cagliari, Italy
\item \Idef{org55}INFN, Sezione di Catania, Catania, Italy
\item \Idef{org56}INFN, Sezione di Padova, Padova, Italy
\item \Idef{org57}INFN, Sezione di Roma, Rome, Italy
\item \Idef{org58}INFN, Sezione di Torino, Turin, Italy
\item \Idef{org59}INFN, Sezione di Trieste, Trieste, Italy
\item \Idef{org60}Inha University, Incheon, Republic of Korea
\item \Idef{org61}Institut de Physique Nucl\'{e}aire d'Orsay (IPNO), Institut National de Physique Nucl\'{e}aire et de Physique des Particules (IN2P3/CNRS), Universit\'{e} de Paris-Sud, Universit\'{e} Paris-Saclay, Orsay, France
\item \Idef{org62}Institute for Nuclear Research, Academy of Sciences, Moscow, Russia
\item \Idef{org63}Institute for Subatomic Physics, Utrecht University/Nikhef, Utrecht, Netherlands
\item \Idef{org64}Institute of Experimental Physics, Slovak Academy of Sciences, Ko\v{s}ice, Slovakia
\item \Idef{org65}Institute of Physics, Homi Bhabha National Institute, Bhubaneswar, India
\item \Idef{org66}Institute of Physics of the Czech Academy of Sciences, Prague, Czech Republic
\item \Idef{org67}Institute of Space Science (ISS), Bucharest, Romania
\item \Idef{org68}Institut f\"{u}r Kernphysik, Johann Wolfgang Goethe-Universit\"{a}t Frankfurt, Frankfurt, Germany
\item \Idef{org69}Instituto de Ciencias Nucleares, Universidad Nacional Aut\'{o}noma de M\'{e}xico, Mexico City, Mexico
\item \Idef{org70}Instituto de F\'{i}sica, Universidade Federal do Rio Grande do Sul (UFRGS), Porto Alegre, Brazil
\item \Idef{org71}Instituto de F\'{\i}sica, Universidad Nacional Aut\'{o}noma de M\'{e}xico, Mexico City, Mexico
\item \Idef{org72}iThemba LABS, National Research Foundation, Somerset West, South Africa
\item \Idef{org73}Jeonbuk National University, Jeonju, Republic of Korea
\item \Idef{org74}Johann-Wolfgang-Goethe Universit\"{a}t Frankfurt Institut f\"{u}r Informatik, Fachbereich Informatik und Mathematik, Frankfurt, Germany
\item \Idef{org75}Joint Institute for Nuclear Research (JINR), Dubna, Russia
\item \Idef{org76}Korea Institute of Science and Technology Information, Daejeon, Republic of Korea
\item \Idef{org77}KTO Karatay University, Konya, Turkey
\item \Idef{org78}Laboratoire de Physique Subatomique et de Cosmologie, Universit\'{e} Grenoble-Alpes, CNRS-IN2P3, Grenoble, France
\item \Idef{org79}Lawrence Berkeley National Laboratory, Berkeley, California, United States
\item \Idef{org80}Lund University Department of Physics, Division of Particle Physics, Lund, Sweden
\item \Idef{org81}Nagasaki Institute of Applied Science, Nagasaki, Japan
\item \Idef{org82}Nara Women{'}s University (NWU), Nara, Japan
\item \Idef{org83}National and Kapodistrian University of Athens, School of Science, Department of Physics , Athens, Greece
\item \Idef{org84}National Centre for Nuclear Research, Warsaw, Poland
\item \Idef{org85}National Institute of Science Education and Research, Homi Bhabha National Institute, Jatni, India
\item \Idef{org86}National Nuclear Research Center, Baku, Azerbaijan
\item \Idef{org87}National Research Centre Kurchatov Institute, Moscow, Russia
\item \Idef{org88}Niels Bohr Institute, University of Copenhagen, Copenhagen, Denmark
\item \Idef{org89}Nikhef, National institute for subatomic physics, Amsterdam, Netherlands
\item \Idef{org90}NRC Kurchatov Institute IHEP, Protvino, Russia
\item \Idef{org91}NRC «Kurchatov Institute»  - ITEP, Moscow, Russia
\item \Idef{org92}NRNU Moscow Engineering Physics Institute, Moscow, Russia
\item \Idef{org93}Nuclear Physics Group, STFC Daresbury Laboratory, Daresbury, United Kingdom
\item \Idef{org94}Nuclear Physics Institute of the Czech Academy of Sciences, \v{R}e\v{z} u Prahy, Czech Republic
\item \Idef{org95}Oak Ridge National Laboratory, Oak Ridge, Tennessee, United States
\item \Idef{org96}Ohio State University, Columbus, Ohio, United States
\item \Idef{org97}Petersburg Nuclear Physics Institute, Gatchina, Russia
\item \Idef{org98}Physics department, Faculty of science, University of Zagreb, Zagreb, Croatia
\item \Idef{org99}Physics Department, Panjab University, Chandigarh, India
\item \Idef{org100}Physics Department, University of Jammu, Jammu, India
\item \Idef{org101}Physics Department, University of Rajasthan, Jaipur, India
\item \Idef{org102}Physikalisches Institut, Eberhard-Karls-Universit\"{a}t T\"{u}bingen, T\"{u}bingen, Germany
\item \Idef{org103}Physikalisches Institut, Ruprecht-Karls-Universit\"{a}t Heidelberg, Heidelberg, Germany
\item \Idef{org104}Physik Department, Technische Universit\"{a}t M\"{u}nchen, Munich, Germany
\item \Idef{org105}Politecnico di Bari, Bari, Italy
\item \Idef{org106}Research Division and ExtreMe Matter Institute EMMI, GSI Helmholtzzentrum f\"ur Schwerionenforschung GmbH, Darmstadt, Germany
\item \Idef{org107}Rudjer Bo\v{s}kovi\'{c} Institute, Zagreb, Croatia
\item \Idef{org108}Russian Federal Nuclear Center (VNIIEF), Sarov, Russia
\item \Idef{org109}Saha Institute of Nuclear Physics, Homi Bhabha National Institute, Kolkata, India
\item \Idef{org110}School of Physics and Astronomy, University of Birmingham, Birmingham, United Kingdom
\item \Idef{org111}Secci\'{o}n F\'{\i}sica, Departamento de Ciencias, Pontificia Universidad Cat\'{o}lica del Per\'{u}, Lima, Peru
\item \Idef{org112}St. Petersburg State University, St. Petersburg, Russia
\item \Idef{org113}Stefan Meyer Institut f\"{u}r Subatomare Physik (SMI), Vienna, Austria
\item \Idef{org114}SUBATECH, IMT Atlantique, Universit\'{e} de Nantes, CNRS-IN2P3, Nantes, France
\item \Idef{org115}Suranaree University of Technology, Nakhon Ratchasima, Thailand
\item \Idef{org116}Technical University of Ko\v{s}ice, Ko\v{s}ice, Slovakia
\item \Idef{org117}Technische Universit\"{a}t M\"{u}nchen, Excellence Cluster 'Universe', Munich, Germany
\item \Idef{org118}The Henryk Niewodniczanski Institute of Nuclear Physics, Polish Academy of Sciences, Cracow, Poland
\item \Idef{org119}The University of Texas at Austin, Austin, Texas, United States
\item \Idef{org120}Universidad Aut\'{o}noma de Sinaloa, Culiac\'{a}n, Mexico
\item \Idef{org121}Universidade de S\~{a}o Paulo (USP), S\~{a}o Paulo, Brazil
\item \Idef{org122}Universidade Estadual de Campinas (UNICAMP), Campinas, Brazil
\item \Idef{org123}Universidade Federal do ABC, Santo Andre, Brazil
\item \Idef{org124}University of Cape Town, Cape Town, South Africa
\item \Idef{org125}University of Houston, Houston, Texas, United States
\item \Idef{org126}University of Jyv\"{a}skyl\"{a}, Jyv\"{a}skyl\"{a}, Finland
\item \Idef{org127}University of Liverpool, Liverpool, United Kingdom
\item \Idef{org128}University of Science and Technology of China, Hefei, China
\item \Idef{org129}University of South-Eastern Norway, Tonsberg, Norway
\item \Idef{org130}University of Tennessee, Knoxville, Tennessee, United States
\item \Idef{org131}University of the Witwatersrand, Johannesburg, South Africa
\item \Idef{org132}University of Tokyo, Tokyo, Japan
\item \Idef{org133}University of Tsukuba, Tsukuba, Japan
\item \Idef{org134}Universit\'{e} Clermont Auvergne, CNRS/IN2P3, LPC, Clermont-Ferrand, France
\item \Idef{org135}Universit\'{e} de Lyon, Universit\'{e} Lyon 1, CNRS/IN2P3, IPN-Lyon, Villeurbanne, Lyon, France
\item \Idef{org136}Universit\'{e} de Strasbourg, CNRS, IPHC UMR 7178, F-67000 Strasbourg, France, Strasbourg, France
\item \Idef{org137}Universit\'{e} Paris-Saclay Centre d'Etudes de Saclay (CEA), IRFU, D\'{e}partment de Physique Nucl\'{e}aire (DPhN), Saclay, France
\item \Idef{org138}Universit\`{a} degli Studi di Foggia, Foggia, Italy
\item \Idef{org139}Universit\`{a} degli Studi di Pavia, Pavia, Italy
\item \Idef{org140}Universit\`{a} di Brescia, Brescia, Italy
\item \Idef{org141}Variable Energy Cyclotron Centre, Homi Bhabha National Institute, Kolkata, India
\item \Idef{org142}Warsaw University of Technology, Warsaw, Poland
\item \Idef{org143}Wayne State University, Detroit, Michigan, United States
\item \Idef{org144}Westf\"{a}lische Wilhelms-Universit\"{a}t M\"{u}nster, Institut f\"{u}r Kernphysik, M\"{u}nster, Germany
\item \Idef{org145}Wigner Research Centre for Physics, Budapest, Hungary
\item \Idef{org146}Yale University, New Haven, Connecticut, United States
\item \Idef{org147}Yonsei University, Seoul, Republic of Korea
\end{Authlist}
\endgroup
\end{document}